%% file: MC3D.tex
\documentclass{egpubl}
\usepackage{eg2022}
 
\ConferenceSubmission   
\usepackage[T1]{fontenc}
\usepackage{dfadobe}  

\biberVersion
\BibtexOrBiblatex
\usepackage[backend=biber,bibstyle=EG,citestyle=alphabetic]{biblatex} 
\electronicVersion
\PrintedOrElectronic
\ifpdf \usepackage[pdftex]{graphicx} \pdfcompresslevel=9
\else \usepackage[dvips]{graphicx} \fi

\usepackage{egweblnk} 
\usepackage[table]{xcolor}
\usepackage{subfig}
\usepackage{wrapfig}
\usepackage{tikz,pgfplots}
\usepackage{adjustbox}
\usepackage{svg}
\usepackage{booktabs}
\usepackage{datatool}
\usepackage{nicefrac}
\usepackage[utf8]{inputenc}
\usepackage[T1]{fontenc}
\usepackage{rotating}
\usepackage{longtable}
\usepackage{csvsimple, longtable, booktabs}

\usepackage{amsthm}
\usepackage{amssymb}
\usepackage{bm}
\usepackage{overpic}
\usepackage{multirow}
\usepackage{comment}
\usepackage{tabto}
\usepackage{xargs}     
\usepackage[ruled,vlined,noend]{algorithm2e}
\usepackage{balance}

\pgfplotsset{compat=1.17}

\definecolor{commentgray}{gray}{0.5}
\definecolor{backgray}{HTML}{f0f0f0}

\SetCommentSty{mycommfont}
\SetAlCapHSkip{1.1mm}

\DeclareMathOperator{\opp}{opp}
\DeclareMathOperator{\alive}{alive}
\DeclareMathOperator{\spli}{iso\_facet}
\DeclareMathOperator{\spliopp}{iso\_facet}
\DeclareMathOperator{\exte}{extent}

\newcommand{\degn}{$90^\circ$}
\newcommand{\dego}{$180^\circ$}
\newcommand{\degt}{$270^\circ$}

\newtheorem{thm}{Theorem}
\newtheorem{definition}[thm]{Definition}

\newcommand\acksname{Acknowledgments}
\specialcomment{acks}{%
  \begingroup
  \section*{\acksname}
  \phantomsection\addcontentsline{toc}{section}{\acksname}
}{%
  \endgroup
}

\title{The 3D Motorcycle Complex for Structured Volume Decomposition\vspace{-0.3cm}}
\author[H. Brückler, O. Gupta, M. Mandad, and M. Campen]{
{\parbox{\textwidth}{\centering\large Hendrik Brückler \quad Ojaswi Gupta \quad Manish Mandad \quad Marcel Campen}\vspace{-0.2cm}}\\%
{\parbox{\textwidth}{\centering Osnabrück University, Germany}}
\vspace{-0.3cm}}

\begin{document}

\input{text/00frontmatter.tex}

\input{text/01introduction.tex}

\input{text/02relatedwork.tex}
\input{text/03background.tex}
\input{text/04motorcyclecomplex.tex}

\input{text/05implementation.tex}

\input{text/06sanitization.tex}
\input{text/07results.tex}
\input{text/08conclusion.tex}

\input{text/99backmatter.tex}

\end{document}


\begin{figure*}[t]
\vspace{-1.0cm}
\textcolor{white}{\rule{\textwidth}{1cm}}
\centering
    \Large\textbf{The 3D Motorcycle Complex for Structured Volume Decomposition\vspace{0.2cm}\\
    --- \;Supplement\; ---}
\end{figure*}

\renewcommand\thetable{S\arabic{table}}
\renewcommand\thefigure{S\arabic{figure}}

\title{The 3D Motorcycle Complex for Structured Volume Decomposition -- Supplementary Material}

\author[H. Brückler, O. Gupta, M. Mandad \& M. Campen]{\parbox{\textwidth}{\centering H. Brückler, O. Gupta, M. Mandad \& M. Campen}}

\renewcommand{\thesection}{\Alph{section}}
\renewcommand{\thefigure}{\Alph{figure}}
\renewcommand{\thetable}{\Alph{table}}
\renewcommand{\thealgocf}{\Alph{algocf}}
\renewcommand{\theequation}{\alph{equation}}

\section{Sparse Serial Motorcycle Complex}

Let us expand on the remark in Sec.~4.2 about the alternative of a serial motorcycle complex construction.
A proposal in \cite[\S7]{eppstein2008motorcycle} for the 2D case is to trace motorcycles in a serial rather than simultaneous manner. While this voids canonicity\footnote{While the simultaneously constructed 2D motorcycle graph yields a \emph{canonical} decomposition, its serial construction voids this property due to order dependence. For 3D neither algorithm yields a canonical partition, as already the crucial 2D right hand arbitration rule does not extend to 3D. The use cases that we discuss do not need this property anyway.}, it enables the option to omit tracing motorcycles that would form removable (though not regular-removable) traces right away. For instance, if the two directions neighboring the next motorcycle's direction around a singularity have been traced (out of or into the singularity) already, the motorcycle can be omitted. The result (called \emph{sparse MC} in the following) will be coarser, though not necessarily irreducible.

Following this idea, Alg.~\ref{alg:meshsparse} is a modified variant of Alg.~1; the parametrization based Alg.~2 can be modified analogously. The key difference is that fire sources (facets incident at singularities) are processed one after the other, and those that are not necessary to establish a valid configuration around a singularity are skipped.

\begin{table}[b]
\vspace{-0.1cm}
\rowcolors{1}{}{backgray}
\begin{filecontents*}{tables/StatsAug05SparseHex.csv}
Model;MCraw;MC;MCsparseRaw;MCsparse;MCrawVsMCsparseRaw;MCvsMCsparseRaw;MCvsMCsparse;MCsparseVsMCsparseRaw
example 3;9087;2877;5125;2691;177.3\%;56.1\%;106.9\%;52.5\%
example 1;3137;1123;1199;962;261.6\%;93.7\%;116.7\%;80.2\%
example 2;233;87;280;101;83.2\%;31.1\%;86.1\%;36.1\%
dragon-hex;979;357;399;317;245.4\%;89.5\%;112.6;79.4\%
gargoyle;720;257;283;220;254.4\%;90.8\%;116.8\%;77.7\%
anc101 a1;1359;460;524;422;259.4\%;87.8\%;109.0\%;80.5\%
fertility-hex;221;76;80;66;276.3\%;95.0\%;115.2\%;82.5\%
pegasus-hex;1035;374;408;287;253.7\%;91.7\%;130.3\%;70.3\%
kiss hex;543;200;244;189;222.5\%;82.0\%;105.8\%;77.5\%
anc101;609;207;136;120;447.8\%;152.2\%;172.5\%;88.2\%
impeller stresstest;184;37;71;61;259.2\%;52.1\%;60.7\%;85.9\%
armadillo hex-a;680;266;292;227;232.9\%;91.1\%;117.2\%;77.7\%
armadillo hex-b;396;147;176;133;225.0\%;83.5\%;110.5\%;75.6\%
;;;\textcolor{black}{\Large$\vphantom{\int^0}${\bfseries\small$\smash[t]{\vdots}$}\vspace{-0.1em}};;;;;
example 5;1;1;1;1;100.0;100.0;100.0\%;100.0\%
\end{filecontents*}
\csvreader[separator=semicolon, before reading=\begin{adjustbox}{max width=\columnwidth}, after reading=\end{adjustbox},tabular=lrrrrrrrr, head to column names=true,
table head =
\bfseries Model & \textcolor{gray}{raw} &  \bfseries MC & \;\;\;\;\;\;\bfseries MC\textsubscript{\large s} & \!\textcolor{gray}{\bfseries MC\textsubscript{\large rs}}  & \;\;\;\;\;\;\;\; \bfseries $\frac{\text{MC}}{\text{MC\textsubscript{\large s}}}$ & \!\!\textcolor{gray}{\bfseries  $\frac{\text{MC}}{\text{MC\textsubscript{\large rs}}}$} \vspace{1mm}\\,
table foot =,
]{tables/StatsAug05SparseHex.csv}{}
{{$\vphantom{\int^0}$\textsc{\Model}\hspace{0.7em}} & \textcolor{gray}{\MCraw} & \!\MC & \!\MCsparseRaw & \!\textcolor{gray}{\MCsparse} & \MCvsMCsparseRaw & \textcolor{gray}{\MCvsMCsparse}}
\caption{Using the dataset from Table 1, reported are the number of blocks in the raw motorcycle complex (raw), fully reduced motorcycle complex (MC), sparse serial motorcycle complex (MC\textsubscript{s}) and its reduced version (MC\textsubscript{rs}) \label{fig:statssparsehex}}
\vspace{-0.1cm}
\end{table}

The condition $\necessary(e,f)$ (line \textbf{1}) is defined as follows. Let $f_{-2}, f_{-1}, f, f_{+1}, f_{+2}$ denote the (possibly cyclically self-overlapping) sequence of facets incident at singular edge $e$ surrounding facet $f$ (in either orientation). Condition $\necessary(e,f)$ is true iff either $f_{-1}$ and $f_{+1}$ are not burnt yet, or $f_{-1}$ and $f_{+2}$ are burnt but not $f_{+1}$, or $f_{-2}$ and $f_{+1}$ are burnt but not $f_{-1}$. In these cases $f$ needs to be burnt (i.e., become part of the motorcycle complex as well) as well---otherwise the complex would contain cells with edges with inner angles of $270^\circ$ (or larger) and would not be a pure cuboid block decomposition.

\vspace{-0.1cm}
\begin{algorithm}
\SetAlgoLined
\DontPrintSemicolon
\ForEach{singular $e$ and $f\!\in\! F_e$}
{\nl \lIf(\tcp*[f]{ignite}){$\necessary(e,f)$}{$Q$.push$((e,f,0))$}
\While{$Q$ non-empty}{
$(e,f,d) \gets Q$.pop()\;
\If(\tcp*[f]{\!\!not\! crossing\! burnt\!\! terrain}){$\alive(e)$}
{tag $f$\tcp*{mark facet as burnt}
\ForEach{regular interior edge $e' \neq e$ incident to $f$}
{\lIf(\tcp*[f]{spread}){$\opp(e',f)$ is not tagged}{$Q$.push$(e', \opp(e',f), d+1)$}}
}}}
\lForEach{boundary facet $f$}{tag $f$}
\caption{Serial Motorcycle Complex of Hex Mesh\label{alg:meshsparse}}
\end{algorithm}
\vspace{-0.1cm}

\begin{table}[b]
\vspace{-0.05cm}
\rowcolors{1}{}{backgray}
\csvreader[respect all, separator=semicolon, before reading=\begin{adjustbox}{max width=\linewidth}, after reading=\end{adjustbox},tabular=lrrrrrr, head to column names=true,
full filter=\ifcsvstrcmp{\Model}{x}
            {\\&&&\huge\bfseries\vphantom{\int^0}\smash[t]{\vdots}&&&&& \csvfilterreject}
                       {\csvfilteraccept},
table head =
\bfseries Model & \textcolor{gray}{raw} &  \bfseries MC & \;\;\;\;\;\;\bfseries MC\textsubscript{\large s} & \!\textcolor{gray}{\bfseries MC\textsubscript{\large rs}}  & \;\;\;\;\;\;\;\; \bfseries $\frac{\text{MC}}{\text{MC\textsubscript{\large s}}}$ & \!\!\textcolor{gray}{\bfseries  $\frac{\text{MC}}{\text{MC\textsubscript{\large rs}}}$} \vspace{1mm}\\,
table foot =,
]{tables/StatsAug05SparseAlgohex.csv}{}
{{$\vphantom{\int^0}$\textsc{\Model}\hspace{3.5em}} & \textcolor{gray}{\MCraw} & \!\MC & \!\MCsparseRaw & \!\textcolor{gray}{\MCsparse} & \MCvsMCsparseRaw\% & \textcolor{gray}{\MCvsMCsparse\%}}
\caption{Using the dataset from Table 2, reported are the number of blocks in the raw motorcycle complex (raw), fully reduced motorcycle complex (MC), sparse serial motorcycle complex  (MC\textsubscript{s}) and its reduced version (MC\textsubscript{rs}) \label{fig:statssparsealgohex}}
\vspace{-0.1cm}
\end{table}

In essence, the algorithm attempts to omit ``every other'' (to the extent permitted by parity) wall around a singularity right away, rather than achieving this via reduction by wall retraction afterwards.
Note that the incorporation of a similar omission strategy directly into the \emph{non-serial} algorithm (as done for the 2D case in \cite[\S3.1]{schertler2018generalized}) would not be straightforward because the a priori omission decision cannot be made simply per singularity in isolation but would require some form of global coordination in the interconnected network of singular arcs in the 3D case.

The result is a block decomposition that can be expected to be coarser than the immediate result (without reduction by wall retraction) of the algorithms from Sec.~5. Indeed this is the case; however, our proposed \emph{reduced} motorcycle complex typically is even simpler than this sparse serial motorcycle complex, as evident from Tables~\ref{fig:statssparsehex} and \ref{fig:statssparsealgohex}. Of course reduction could also be applied to the sparse MC, but this yields no consistent benefit (last column).

\section{Constraint System Simplification}
\label{app:trulyseamless}

We show that, analogous to \cite[\S5.3]{mandad2019exact}, the 3D constraint system can be transformed such that only a small subsystem needs to be considered in the exact solver; all other variables are deduced in a back-substitution manner.
The subsystem contains only the variables associated with node vertices. This shows that the small system set up in Sec.~6.2.2 is indeed sufficient.

At vertices not incident to any cut or align facets, incident tetrahedra form a single sector and hence, share parametrization values. The equations corresponding to edges incident to such vertices are trivially satisfied and can be left out of the system. This leaves us with a system over the variables of vertices on sheets.

\paragraph*{Cut Sheets.}
For each sheet
we can denote the parametrization variables of a vertex $p$ in the two sectors on the two sides of the sheet as $\bm{u}^+_p$ and $\bm{u}^-_p$.
Given a sheet with transition function $\pi$, the transition constraint for an edge $ab$ on the sheet then has the form
$
\bm{u}^+_b - \bm{u}^+_a = \pi (\bm{u}^-_b - \bm{u}^-_a).
$

Given a sequence of vertices $\bm{u}^\pm_0, \bm{u}^\pm_1, \ldots, \bm{u}^\pm_n$ forming a chain of edges on a sheet, the equation corresponding to the $k$-th edge is:
$
\tau^{w}_{k}:\; \bm{u}^+_k - \bm{u}^+_{k-1} = \pi (\bm{u}^-_k - \bm{u}^-_{k-1}).
$

Cumulative sums of these equations have a simple form, namely
\vspace{-0.45cm}
\begin{equation}
\label{eq:cumulative_sum}
\sum\limits_{i=1}^k \tau^{w}_{i}:\; \bm{u}^+_k - \bm{u}^+_0 = \pi (\bm{u}^-_k - \bm{u}^-_0),
\end{equation}
\vspace{-0.4cm}

All transition constraints of a sheet can hence be rewritten with respect to one selected \emph{base} node vertex of the sheet, with variable $\bm{u}_0$. Re-ordering the variables in the system such that those corresponding to node sectors come last, the following structure is obtained per sheet:
\begin{samepage}
\quad\vspace{-0.35cm}\quad
$$\hspace{5.4cm}\;\;\;\;\;\bm{u}^{-}_{0} \:\:\:\:\: \bm{u}^{+}_{0}$$
\vspace{-.5cm}
\begin{equation}
\label{eq:sys:A1}
\left[\begin{array}{rrrrrr|rrrrrrrr}
\bm{1}&&&\!\!\!\!\!\!\!\!-\pi &&&&&&&&&-\bm{1}&\pi\\
&\!\!\!\ddots&&&\!\!\!\!\!\!\!\!\ddots&&&&&&&&\multicolumn{2}{c}{\vdots}\\
&&\bm{1}&&&\!\!\!\!\!\!\!\!-\pi &&&&&&&-\bm{1}&\pi\\
&&&&&&\bm{1}&&&\!\!\!\!\!\!\!\!\!\!\!-\pi &&&-\bm{1}&\pi\\
&&&&&&&\!\!\!\ddots&&&\!\!\!\!\!\!\!\!\ddots&&\multicolumn{2}{c}{\vdots}\\
&&&&&&&&\bm{1}&&&\!\!\!\!\!\!\!\!-\pi &-\bm{1}&\pi
\end{array}\right]
\end{equation}
\end{samepage}
where each entry is a $3\times 3$ block ($\pi$: rotational matrix of transition; $\bm{1} = \text{diag}(1,1,1)$), because $\bm u$ has three components. The vertical bar separates non-node (left) from node vertex variables (right).

\paragraph*{Align Sheets.} Analogously, cumulative sums of alignment constraints can be built, forming this structure per align sheet:
$$\hspace{3.8cm} \bm{u}_{0}$$
\vspace{-0.5cm}
\begin{equation}
\label{eq:sys:Ax}
\left[\begin{array}{rrr|rrrr}
\bm{1}^k&&&&&&\!\!\!\!-\bm{1}^k\\
&\!\!\!\ddots&&&&&\vdots\\
&&\bm{1}^k&&&&\!\!\!\!-\bm{1}^k\\
&&&\bm{1}^k&&&\!\!\!\!-\bm{1}^k\\
&&&&\!\!\!\ddots&&\vdots\\
&&&&&\bm{1}^k&\!\!\!\!-\bm{1}^k
\end{array}\right]
\end{equation}
where $\bm{1}^k$ is a $1\times 3$ block: $[1,0,0]$ for $k=0$, $[0,1,0]$ for $k=1$, $[0,0,1]$ for $k=2$, where $k$ is the aligned coordinate component.

\paragraph*{Global System} Combining these systems over all sheets will yield the global system. However, 
not all non-node vertex variables appear in only one of these systems. It is therefore not evident that the combined system maintains an upper triangular structure among the non-node vertices (the left parts in the above matrices).
Concretely, among the non-node vertex variables, those on branches
appear in two equations each, because each sector incident on an inner branch vertex is bounded by two sheets. We show that the union of these equations still forms an upper triangular system.

\begin{wrapfigure}[6]{r}{0.23\linewidth}
\vspace{-6.5mm}
\hspace{-0.6cm}
\begin{overpic}[width=1.1\linewidth]{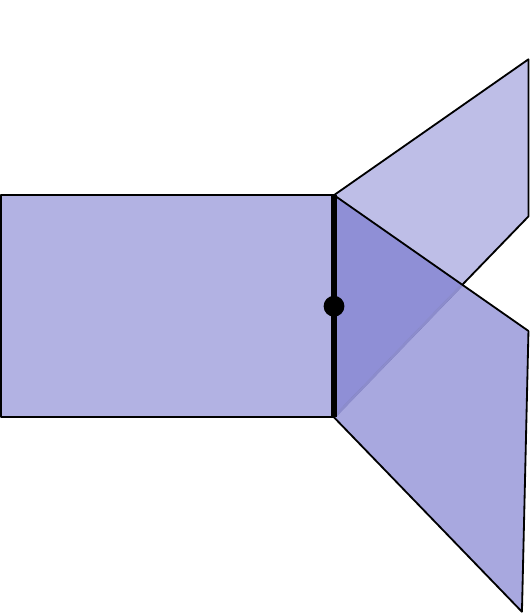}
 \put(9,46){$\pi_{1}$}
 \put(67,26){$\pi_{2}$}
 \put(67,68.3){$\pi_{3}$}
 \put(39,79){$\bm{p_{0}}$}
 \put(39,20){$\bm{p_{1}}$}
 \put(85.9,53.8){$\bm{p_{2}}$}
\end{overpic}
\end{wrapfigure}
Given a cycle of $n$ sheets incident on a branch with transition functions $\pi_{1}, \pi_{2}\ldots \pi_{n}$ and base node sector variables $\bm{u}^\pm_1, \bm{u}^\pm_2\ldots \bm{u}^\pm_{n}$, respectively, the transition equation for a vertex with sector variables $\bm{p}_0, \bm{p}_1\ldots \bm{p}_{n-1}$ (see inset figure for n = 3) corresponding to the $k$-th sheet will be:
$\bm{p}_{k} - \bm{u}^+_k = \pi_k (\bm{p}_{k-1} - \bm{u}^-_k)$.

Via variable elimination and reordering these can be turned into an upper triangular structure, namely:
\vspace{-0.3cm}

\begin{samepage}
\vspace{-0.2cm}
$$
\!\!\!\!\!\!\!\bm{p}_{1}\:\: \bm{p}_{2} \qquad\qquad \bm{p}_{0} \qquad\:\: \bm{u}^-_{1} \qquad\:\:\: \bm{u}^+_{1} \qquad\qquad\quad \bm{u}^-_{n} \:\:\:\: \bm{u}^+_{n}\quad
$$
\vspace{-0.4cm}
\begin{equation}
\label{eq:sys:A2}
\!\left[\begin{array}{rrrr|rrrr}
\!\!\bm{1}&&&-\pi_{1}& \pi_{1}& - \bm{1}&\\

&\!\!\!\!\!\!\bm{1}&&-\pi_{2}\pi_{1}&\pi_{2}\pi_{1}& -\pi_{2}&\pi_{2}\: - \bm{1}&\\
&&\!\!\!\!\!\!\!\!\ddots &\vdots&\vdots&\vdots& \ddots\\

&&&\!\!\!\!\!\!\!\!\bm{1}-\pi_{n}{\scriptscriptstyle{\ldots}}\pi_{1}&\pi_{n}{\scriptscriptstyle{\ldots}}\pi_{1} &\!\!\!\! -\pi_{n}{\scriptscriptstyle{\ldots}}\pi_{2}&\ldots& \pi_{n}\:-\bm{1}\\
\end{array}\right]\!\!
\end{equation}
\end{samepage}

In case of a branch lying on the boundary, the sheets around it can be arranged such that the first and last are alignment sheets, i.e., $a_1, \bm{u}^\pm_1, \bm{u}^\pm_2\ldots \bm{u}^\pm_{n-2}, a_2$. This yields a lower bidiagonal structure for the non-node variables (easily transformable into triangular form):\vspace{-0.3cm}

\begin{samepage}
\begin{equation}
\label{eq:sys:A3}
\left[\begin{array}{rrrr|rrrrr}
\bm{1}^{a_1}&&&& &&&\!\!\!\!\!\!\bm{1}^{a_1}\\

\!\!-\pi_{1}&\!\!\!\!\bm{1}&&& \pi_{1}\: - \bm{1}&\\
&\!\!\ddots&\!\!\ddots &&& \!\!\!\!\!\!\!\!\!\ddots\!\!\!\\
&&\!\!\!\!-\pi_{n-2}& \!\bm{1}\:\:\:\: & && \!\!\!\!\pi_{n-2}\:-\bm{1} \\

&&&\bm{1}^{a_2}& &&&&\!\!\!\!\!\!\!\!\bm{1}^{a_2}\!\!\!
\end{array}\right]\!\!
\end{equation}
\end{samepage}

Globally ordering all variables in the system such that those corresponding to node sectors come last, the cumulative transition and alignment equations form the following constraint system:
\vspace{-0.1cm}

\begin{samepage}

$$\left[\begin{array}{ccc|r}

A_1 &&& \\
&\!\!\!\!\!\!\!\! A_2\! && \;\;B\;\; \\
&&\!\!\!\!\!\!\!\!\!\!\!  A_3\!\!  & \\

\hline
\vspace{-0.0cm}

&\!\!\!\!\!\!\!\!\!\!
\begin{matrix}
 & \!\!\!\! & \vspace{-0.15cm}\\
 & \!0 & \!\!\!\!\vspace{-0.25cm}\\
 & \!\!\!\! & 
\end{matrix}\!\!\!\! && \;\;C\;\;

\end{array}\right]$$

\end{samepage}

\vspace{0.3cm}

\noindent where all the sub matrices $A_i$ are upper triangular (formed by the left parts of the above matrices), containing constraints corresponding to, respectively, non-branch sheet vertices \eqref{eq:sys:A1}+\eqref{eq:sys:Ax}, non-node interior branch vertices \eqref{eq:sys:A2}, and non-node boundary branch vertices \eqref{eq:sys:A3}. The small block $C$
is exactly the system we set up and solve in Sec.~6.2.2, and the described simple subsequent propagation process corresponds to back-substitution through $B$ and $A_i$ to the non-node vertex variables.

\begin{table*}[h!]
\section{Complete Version of Table 1}\vspace{0.4cm}

\rowcolors{1}{}{backgray}
\begin{filecontents*}{tables/StatsMay18part1.csv}
Model;hexMeshValid;BC;BCR;MCraw;MCplus;MCpluspercent;Tpercent;MC;MCpercent
2018 - Fuzzy clustering based pseudo-swept volume decomposition for hexahedral meshing_Example_3;yes;406136;67828;9087;5780;1.42;41.63;2877;0.71
2018 - Fuzzy clustering based pseudo-swept volume decomposition for hexahedral meshing_Example_1;yes;74331;11385;3137;2248;3.02;15.09;1123;1.51
2018 - Fuzzy clustering based pseudo-swept volume decomposition for hexahedral meshing_Example_2;yes;3253;678;233;195;5.99;13.67;87;2.67
2016 - All-Hex Meshing Using Closed-Form Induced Polycube_dragon-hex;yes;12488;2959;979;724;5.8;36.33;357;2.86
2014 - l1-Based Construction of Polycube Maps from Complex Shapes_gargoyle;yes;7563;1967;720;546;7.22;37.62;257;3.4
2011 - All-Hex Mesh Generation via Volumetric PolyCube Deformation_anc101_a1;yes;12336;3118;1359;846;6.86;45.33;460;3.73
2016 - All-Hex Meshing Using Closed-Form Induced Polycube_fertility-hex;yes;2002;548;221;189;9.44;27.02;76;3.8
2016 - All-Hex Meshing Using Closed-Form Induced Polycube_pegasus-hex;yes;9745;2415;1035;729;7.48;36.29;374;3.84
2016 - Efficient Volumetric PolyCube-Map Construction_Kiss_hex;yes;5019;1194;543;385;7.67;39.3;200;3.98
2014 - l1-Based Construction of Polycube Maps from Complex Shapes_anc101;yes;5009;1283;609;347;6.93;38.9;207;4.13
2015 - Practical Hex-Mesh Optimization via Edge-Cone Rectification_impeller_stresstest_in;yes;878;176;184;124;14.12;23.86;37;4.21
2015 - Practical Hex-Mesh Optimization via Edge-Cone Rectification_impeller_stresstest_out;yes;878;176;184;124;14.12;23.86;37;4.21
2012 - All-Hex Meshing using Singularity-Restricted Field_impeller;yes;878;182;184;124;14.12;23.86;39;4.44
2016 - Efficient Volumetric PolyCube-Map Construction_Armadillo_hex-a;yes;5960;1491;680;516;8.66;34.19;266;4.46
2016 - Efficient Volumetric PolyCube-Map Construction_Armadillo_hex-b;yes;3265;820;396;296;9.07;31.45;147;4.5
2014 - l1-Based Construction of Polycube Maps from Complex Shapes_dancing-children-2;yes;5482;1516;672;533;9.72;32.9;259;4.72
2015 - Practical Hex-Mesh Optimization via Edge-Cone Rectification_dancingchildren_in;yes;5482;1406;703;546;9.96;34.52;274;5
2015 - Practical Hex-Mesh Optimization via Edge-Cone Rectification_dancingchildren_out;yes;5482;1406;703;546;9.96;34.52;274;5
2016 - All-Hex Meshing Using Closed-Form Induced Polycube_chinese-lion-hex;yes;6235;1468;818;589;9.45;38.82;321;5.15
2016 - Efficient Volumetric PolyCube-Map Construction_Dancing_Children_hex;yes;4755;1181;682;532;11.19;31.97;247;5.19
2012 - All-Hex Meshing using Singularity-Restricted Field_fertility;yes;1352;342;234;188;13.91;29.45;72;5.33
2015 - Practical Hex-Mesh Optimization via Edge-Cone Rectification_armadillo_in;yes;2112;348;308;265;12.55;24.53;114;5.4
2015 - Practical Hex-Mesh Optimization via Edge-Cone Rectification_armadillo_out;yes;2112;348;308;265;12.55;24.53;114;5.4
2015 - Practical Hex-Mesh Optimization via Edge-Cone Rectification_armadillo_stresstest_in;yes;2112;348;308;265;12.55;24.53;114;5.4
2015 - Practical Hex-Mesh Optimization via Edge-Cone Rectification_armadillo_stresstest_out;yes;2112;348;308;265;12.55;24.53;114;5.4
2014 - l1-Based Construction of Polycube Maps from Complex Shapes_cognit;yes;5194;1350;759;574;11.05;32.35;298;5.74
2016 - Efficient Volumetric PolyCube-Map Construction_Dragon_hex;yes;3655;853;569;442;12.09;28.81;212;5.8
2014 - l1-Based Construction of Polycube Maps from Complex Shapes_elephant;yes;2842;692;415;346;12.17;28.97;167;5.88
2016 - Efficient Volumetric PolyCube-Map Construction_Elephant_hex;yes;3105;770;527;399;12.85;29.15;189;6.09
2016 - Efficient Volumetric PolyCube-Map Construction_Bunny_hex;yes;1282;348;257;187;14.59;30.74;80;6.24
2011 - All-Hex Mesh Generation via Volumetric PolyCube Deformation_Kiss_hex_coarse;yes;3690;1023;557;402;10.89;38.44;231;6.26
2015 - Practical Hex-Mesh Optimization via Edge-Cone Rectification_dragon_in;yes;2019;495;340;220;10.9;27.43;127;6.29
2015 - Practical Hex-Mesh Optimization via Edge-Cone Rectification_dragon_out;yes;2019;495;340;220;10.9;27.43;127;6.29
2011 - All-Hex Mesh Generation via Volumetric PolyCube Deformation_casting;yes;2805;701;524;409;14.58;29.47;185;6.6
2011 - All-Hex Mesh Generation via Volumetric PolyCube Deformation_kiss_hex;yes;3690;1075;688;434;11.76;39.25;255;6.91
2016 - Polycube simplification for coarse layouts of surfaces and volumes_chinese_dragon_polycube_in;yes;810;189;174;147;18.15;22.66;56;6.91
2016 - All-Hex Meshing Using Closed-Form Induced Polycube_grayloc-hex;yes;3183;804;604;477;14.99;27.97;222;6.97
2013 - PolyCut - Monotone Graph-Cuts for PolyCube Base-Complex Construction_carter_hex_opt;yes;2500;635;537;410;16.4;33.74;182;7.28
2011 - All-Hex Mesh Generation via Volumetric PolyCube Deformation_Stab3_refine3;yes;2227;580;419;290;13.02;36.08;163;7.32
2011 - All-Hex Mesh Generation via Volumetric PolyCube Deformation_bunny_hex;yes;1324;373;275;194;14.65;31.87;97;7.33
2016 - Polycube simplification for coarse layouts of surfaces and volumes_rockerarm_model_in;yes;678;152;162;130;19.17;25.86;50;7.37
2016 - Polycube simplification for coarse layouts of surfaces and volumes_rockerarm_polycube_in;yes;678;152;162;130;19.17;25.86;50;7.37
2017 - A global approach to multi-axis swept mesh generation_Example_4;yes;1360;324;302;201;14.78;24.19;101;7.43
2014 - l1-Based Construction of Polycube Maps from Complex Shapes_dragon;yes;3977;998;836;661;16.62;28.78;301;7.57
2016 - All-Hex Meshing Using Closed-Form Induced Polycube_rockerarm-hex;yes;1202;306;223;181;15.06;28.03;91;7.57
2017 - A global approach to multi-axis swept mesh generation_Example_1;yes;1037;243;225;150;14.46;22.32;81;7.81
2016 - Efficient Volumetric PolyCube-Map Construction_Buste_hex;yes;1081;275;244;182;16.84;32.4;85;7.86
2016 - Efficient Volumetric PolyCube-Map Construction_Greek_Sculpture_hex;yes;1476;315;310;244;16.53;28.44;118;7.99
2019 - Singularity Structure Simplification of Hexahedral Mesh via Weighted Ranking_deckel_input;yes;53116;12557;10922;8823;16.61;33.39;4249;8
2016 - Polycube simplification for coarse layouts of surfaces and volumes_bunny_model_in;yes;637;141;151;124;19.47;26.89;51;8.01
2016 - Polycube simplification for coarse layouts of surfaces and volumes_bunny_polycube_in;yes;637;141;151;124;19.47;26.89;51;8.01
2016 - Efficient Volumetric PolyCube-Map Construction_Carter_hex;yes;2788;616;630;450;16.14;34.55;228;8.18
2013 - PolyCut - Monotone Graph-Cuts for PolyCube Base-Complex Construction_BU_hex_opt;yes;580;141;132;113;19.48;27.21;48;8.28
2014 - l1-Based Construction of Polycube Maps from Complex Shapes_rockerArm_2;yes;835;167;189;149;17.84;30.57;71;8.5
2016 - Efficient Volumetric PolyCube-Map Construction_Bimba_hex-c;yes;760;176;204;146;19.21;30.68;68;8.95
2012 - All-Hex Meshing using Singularity-Restricted Field_rockerarm;yes;578;135;150;120;20.76;30.66;52;9
2011 - All-Hex Mesh Generation via Volumetric PolyCube Deformation_bumpy_torus;yes;2518;631;595;446;17.71;32.19;228;9.05
2011 - All-Hex Mesh Generation via Volumetric PolyCube Deformation_BU_remesh_hex;yes;1098;308;301;202;18.4;33.15;102;9.29
2017 - A global approach to multi-axis swept mesh generation_Example_3;yes;1122;280;341;264;23.53;18.57;105;9.36
2016 - Efficient Volumetric PolyCube-Map Construction_Sphinx_hex-e;yes;801;229;222;161;20.1;32.19;75;9.36
2016 - Efficient Volumetric PolyCube-Map Construction_Rocker_hex;yes;730;187;199;156;21.37;28.02;69;9.45
2017 - A global approach to multi-axis swept mesh generation_Example_2;yes;974;224;288;202;20.74;24.89;95;9.75
2016 - Polycube simplification for coarse layouts of surfaces and volumes_teapot_model_in;yes;328;68;108;89;27.13;28.6;32;9.76
2016 - Polycube simplification for coarse layouts of surfaces and volumes_teapot_polycube_in;yes;328;68;108;89;27.13;28.6;32;9.76
2014 - l1-Based Construction of Polycube Maps from Complex Shapes_bumpy_torus;yes;2254;640;592;488;21.65;29.29;224;9.94
2013 - PolyCut - Monotone Graph-Cuts for PolyCube Base-Complex Construction_rocker_arm_hex_opt;yes;664;176;184;138;20.78;28.81;67;10.09
2016 - All-Hex Meshing Using Closed-Form Induced Polycube_kitten-hex;yes;208;57;71;57;27.4;24.66;21;10.1
2016 - Skeleton-driven Adaptive Hexahedral Meshing of Tubular Shapes_warrior_graded;yes;4869;1306;1447;1282;26.33;30.81;492;10.1
2019 - Dual Sheet Meshing - An Interactive Approach to Robust Hexahedralization_fandisk.liu18;yes;89;20;35;26;29.21;13.66;9;10.11
2014 - l1-Based Construction of Polycube Maps from Complex Shapes_angel_1;yes;1284;336;362;266;20.72;34.99;131;10.2
2016 - All-Hex Meshing Using Closed-Form Induced Polycube_carter-hex;yes;1101;235;359;266;24.16;35.2;114;10.35
2016 - Skeleton-driven Adaptive Hexahedral Meshing of Tubular Shapes_dinopet_graded;yes;2253;617;721;630;27.96;32.67;234;10.39
2015 - Practical Hex-Mesh Optimization via Edge-Cone Rectification_bust_in;yes;494;102;128;110;22.27;27.34;52;10.53
2015 - Practical Hex-Mesh Optimization via Edge-Cone Rectification_bust_out;yes;494;102;128;110;22.27;27.34;52;10.53
2015 - Practical Hex-Mesh Optimization via Edge-Cone Rectification_KingKong_in;yes;180;45;71;62;34.44;25;19;10.56
2015 - Practical Hex-Mesh Optimization via Edge-Cone Rectification_KingKong_out;yes;180;45;71;62;34.44;25;19;10.56
2014 - l1-Based Construction of Polycube Maps from Complex Shapes_kiss;yes;1896;567;560;418;22.05;32.88;205;10.81
2016 - Skeleton-driven Adaptive Hexahedral Meshing of Tubular Shapes_dino_graded;yes;904;275;306;261;28.87;33.99;99;10.95
2016 - Structured Volume Decomposition via Generalized Sweeping_fertility_1;yes;301;73;119;105;34.88;22.71;33;10.96
2019 - Selective Padding for Polycube‐Based Hexahedral Meshing_Double_hinge_WH;yes;153;16;43;43;28.1;10;17;11.11
2016 - Structured Volume Decomposition via Generalized Sweeping_fertility_2;yes;301;50;119;105;34.88;22.71;34;11.3
2016 - Structured Volume Decomposition via Generalized Sweeping_fertility_3;yes;301;71;119;105;34.88;22.71;34;11.3
2013 - PolyCut - Monotone Graph-Cuts for PolyCube Base-Complex Construction_bunny_hex_opt;yes;580;120;170;127;21.9;30.17;66;11.38
2016 - Polycube simplification for coarse layouts of surfaces and volumes_asm_model_out;yes;122;33;68;64;52.46;14.29;14;11.48
2016 - Polycube simplification for coarse layouts of surfaces and volumes_asm_polycube_out;yes;122;33;68;64;52.46;14.29;14;11.48
2017 - Hexahedral Mesh Generation via Constrained Quadrilateralization_joint;yes;83;18;41;27;32.53;15.43;10;12.05
2016 - Polycube simplification for coarse layouts of surfaces and volumes_fandisk_polycube_in;yes;229;49;89;70;30.57;21.43;29;12.66
\end{filecontents*}
\csvreader[respect all, separator=semicolon, before reading=\begin{adjustbox}{max width=\linewidth}, after reading=\end{adjustbox},tabular=lrrrrrrrr, head to column names=true,
table head =
\bfseries Model & \bfseries BC &  \textcolor{gray}{BC\textsuperscript{--}} & \;\;\;\;\;\;\;\;\;\textcolor{gray}{raw} & \bfseries \!MC\textsuperscript{+}\!\! & \;\;\;\;\;\;\bfseries $\frac{\text{MC\textsuperscript{+}}}{\text{BC}}$ & \textcolor{gray}{\;\;\;\;\;\;\;\;\;\;\;\;\;T\;\;} & \;\;\;\;\;\;\bfseries MC & \!\!\bfseries $\frac{\text{MC}}{\text{BC}}$ \vspace{1mm}\\,
table foot =,
]{tables/StatsMay18part1.csv}{}
{{\textsc{\Model}}\;\;\;\;\;\;\;\; & \BC & \textcolor{gray}{\BCR} & \textcolor{gray}{\MCraw} & \MCplus & \!\MCpluspercent\% & \textcolor{gray}{\Tpercent\%} & \MC & \!\MCpercent\%}
\vspace{-0.1cm}
\caption{Statistics on a dataset of
{hexahedral meshes}. Numbers of blocks in the base complex (BC), reduced base complex (BC\textsuperscript{--}), raw motorcycle complex (raw), reduced motorcycle complex with preserved singularity-adjacent walls (MC\textsuperscript{+}), fully reduced motorcycle complex (MC), percentage of arcs that are T-arcs (T).\label{fig:stats}}
\vspace{-0.45cm}
\end{table*}

\addtocounter{table}{-1}    

\begin{table*}[h!]
\rowcolors{1}{}{backgray}
\begin{filecontents*}{tables/StatsMay18part2.csv}
Model;hexMeshValid;BC;BCR;MCraw;MCplus;MCpluspercent;Tpercent;MC;MCpercent
2014 - l1-Based Construction of Polycube Maps from Complex Shapes_rockerArm_1;yes;686;183;247;186;27.11;27.02;87;12.68
2015 - Practical Hex-Mesh Optimization via Edge-Cone Rectification_cap_in;yes;327;97;124;92;28.13;34.55;42;12.84
2015 - Practical Hex-Mesh Optimization via Edge-Cone Rectification_cap_out;yes;327;97;124;92;28.13;34.55;42;12.84
2016 - Polycube simplification for coarse layouts of surfaces and volumes_rockerarm_model_out;yes;316;62;130;116;36.71;16.14;41;12.97
2013 - PolyCut - Monotone Graph-Cuts for PolyCube Base-Complex Construction_fertility_hex_opt;yes;693;153;243;202;29.15;25.95;91;13.13
2016 - Polycube simplification for coarse layouts of surfaces and volumes_rockerarm_polycube_out;yes;348;60;132;117;33.62;19.5;46;13.22
2016 - Efficient Volumetric PolyCube-Map Construction_Bimba_hex-d;yes;196;60;97;63;32.14;23.77;26;13.27
2019 - Selective Padding for Polycube‐Based Hexahedral Meshing_Lego_L2;yes;525;83;221;167;31.81;29.6;70;13.33
2019 - Dual Sheet Meshing - An Interactive Approach to Robust Hexahedralization_fandisk;yes;89;14;36;27;30.34;10.99;12;13.48
2016 - Efficient Volumetric PolyCube-Map Construction_Blade_hex;yes;389;90;142;130;33.42;20.78;53;13.62
2016 - Efficient Volumetric PolyCube-Map Construction_Fertility_hex-largel;yes;633;148;244;211;33.33;22.91;87;13.74
2016 - Efficient Volumetric PolyCube-Map Construction_Fertility_hex-small;yes;621;158;243;210;33.82;23.37;87;14.01
2019 - Singularity Structure Simplification of Hexahedral Mesh via Weighted Ranking_fertility_input;yes;20840;5523;8344;7105;34.09;30.4;2951;14.16
2014 - l1-Based Construction of Polycube Maps from Complex Shapes_angel_2;yes;302;75;116;103;34.11;23.08;43;14.24
2019 - Selective Padding for Polycube‐Based Hexahedral Meshing_Double_hinge_NH;yes;105;15;35;35;33.33;6.67;15;14.29
2016 - Polycube simplification for coarse layouts of surfaces and volumes_table1_polycube_in;yes;195;42;93;80;41.03;25.17;28;14.36
2011 - All-Hex Mesh Generation via Volumetric PolyCube Deformation_fertility_refine;yes;598;161;243;209;34.95;24.86;87;14.55
2014 - l1-Based Construction of Polycube Maps from Complex Shapes_canewt;yes;879;243;357;270;30.72;30.87;128;14.56
2019 - Dual Sheet Meshing - An Interactive Approach to Robust Hexahedralization_hanger;yes;41;11;30;20;48.78;15.33;6;14.63
2019 - Singularity Structure Simplification of Hexahedral Mesh via Weighted Ranking_buste_input;yes;18355;4925;7667;6570;35.79;30.83;2722;14.83
2012 - All-Hex Meshing using Singularity-Restricted Field_bunny;yes;259;52;101;93;35.91;19.65;39;15.06
2019 - Selective Padding for Polycube‐Based Hexahedral Meshing_Joint;yes;119;17;42;33;27.73;12.92;18;15.13
2012 - All-Hex Meshing using Singularity-Restricted Field_rod;yes;66;9;37;27;40.91;18.09;10;15.15
2019 - Dual Sheet Meshing - An Interactive Approach to Robust Hexahedralization_joint;yes;59;10;33;24;40.68;10.37;9;15.25
2016 - All-Hex Meshing Using Closed-Form Induced Polycube_hollow-eight-hex;yes;249;61;108;76;30.52;39.2;38;15.26
2019 - Singularity Structure Simplification of Hexahedral Mesh via Weighted Ranking_buste_output;yes;190;45;90;82;43.16;17.75;29;15.26
2016 - Structured Volume Decomposition via Generalized Sweeping_cat_2_padded;yes;19;3;19;14;73.68;23.29;3;15.79
2016 - Structured Volume Decomposition via Generalized Sweeping_Dolphin_padded;yes;19;3;19;14;73.68;23.29;3;15.79
2016 - Structured Volume Decomposition via Generalized Sweeping_femur1_padded;yes;19;3;19;14;73.68;23.29;3;15.79
2016 - Structured Volume Decomposition via Generalized Sweeping_rabbit_padded;yes;19;3;19;14;73.68;23.29;3;15.79
2016 - Structured Volume Decomposition via Generalized Sweeping_rotellipse_padded;yes;19;3;19;14;73.68;23.29;3;15.79
2016 - Polycube simplification for coarse layouts of surfaces and volumes_asm_model_in;yes;200;52;80;66;33;23.08;32;16
2016 - Polycube simplification for coarse layouts of surfaces and volumes_asm_polycube_in;yes;200;52;80;66;33;23.08;32;16
2019 - Selective Padding for Polycube‐Based Hexahedral Meshing_Bearing;yes;184;46;70;64;34.78;12.73;30;16.3
2016 - Polycube simplification for coarse layouts of surfaces and volumes_femur_model_out;yes;110;29;57;55;50;16.48;18;16.36
2014 - l1-Based Construction of Polycube Maps from Complex Shapes_fertility;yes;460;124;212;188;40.87;20.33;76;16.52
2019 - Singularity Structure Simplification of Hexahedral Mesh via Weighted Ranking_pig_output;yes;876;202;420;339;38.7;28.33;146;16.67
2016 - All-Hex Meshing Using Closed-Form Induced Polycube_joint-hex;yes;59;10;31;22;37.29;15.72;10;16.95
2016 - Polycube simplification for coarse layouts of surfaces and volumes_cubespikes_model_in;yes;276;53;123;105;38.04;21.02;47;17.03
2016 - Polycube simplification for coarse layouts of surfaces and volumes_cubespikes_polycube_in;yes;276;53;123;105;38.04;21.02;47;17.03
2012 - All-Hex Meshing using Singularity-Restricted Field_ellipsoid-A;yes;34;7;27;14;41.18;32.18;6;17.65
2019 - Singularity Structure Simplification of Hexahedral Mesh via Weighted Ranking_bottle2_output;yes;318;73;175;157;49.37;18.26;57;17.92
2012 - All-Hex Meshing using Singularity-Restricted Field_hanger;yes;50;7;27;22;44;16.08;9;18
2012 - All-Hex Meshing using Singularity-Restricted Field_joint;yes;83;21;47;28;33.73;17.8;15;18.07
2017 - Hexahedral Mesh Generation via Constrained Quadrilateralization_pone.0177603.s003;yes;83;21;47;28;33.73;17.8;15;18.07
2019 - Singularity Structure Simplification of Hexahedral Mesh via Weighted Ranking_toy2_input;yes;14288;3857;7161;6153;43.06;30.76;2585;18.09
2016 - Polycube simplification for coarse layouts of surfaces and volumes_table1_polycube_out;yes;149;31;86;72;48.32;25.77;27;18.12
2016 - Polycube simplification for coarse layouts of surfaces and volumes_fandisk_polycube_out;yes;160;46;79;67;41.88;18.01;29;18.13
2019 - Singularity Structure Simplification of Hexahedral Mesh via Weighted Ranking_toy1_input;yes;18883;5246;9461;8122;43.01;29.98;3427;18.15
2016 - Polycube simplification for coarse layouts of surfaces and volumes_femur_polycube_out;yes;110;30;57;55;50;16.48;20;18.18
2016 - Polycube simplification for coarse layouts of surfaces and volumes_bunny_model_out;yes;197;53;109;99;50.25;13.94;36;18.27
2016 - Polycube simplification for coarse layouts of surfaces and volumes_bunny_polycube_out;yes;197;53;109;99;50.25;13.94;36;18.27
2012 - All-Hex Meshing using Singularity-Restricted Field_fandisk;yes;49;11;25;22;44.9;11.61;9;18.37
2017 - Hexahedral Mesh Generation via Constrained Quadrilateralization_fandisk;yes;49;11;28;22;44.9;11.61;9;18.37
2017 - Hexahedral Mesh Generation via Constrained Quadrilateralization_pone.0177603.s002;yes;49;16;27;22;44.9;11.61;9;18.37
2019 - Dual Sheet Meshing - An Interactive Approach to Robust Hexahedralization_rockarm;yes;119;35;59;41;34.45;18.12;22;18.49
2016 - Polycube simplification for coarse layouts of surfaces and volumes_hand_model_in;yes;172;35;81;77;44.77;17.77;32;18.6
2016 - Polycube simplification for coarse layouts of surfaces and volumes_hand_polycube_in;yes;172;35;81;77;44.77;17.77;32;18.6
2019 - Singularity Structure Simplification of Hexahedral Mesh via Weighted Ranking_pig_input;yes;13987;3768;7144;6078;43.45;30.98;2610;18.66
2019 - Singularity Structure Simplification of Hexahedral Mesh via Weighted Ranking_bottle2_input;yes;35860;9968;19360;16349;45.59;31.63;6816;19.01
2016 - Structured Volume Decomposition via Generalized Sweeping_kitten_with_bifur_2;yes;26;8;22;22;84.62;10.53;5;19.23
2019 - Singularity Structure Simplification of Hexahedral Mesh via Weighted Ranking_deckel_output;yes;806;237;450;382;47.39;24.24;155;19.23
2016 - Polycube simplification for coarse layouts of surfaces and volumes_femur_model_in;yes;145;36;63;59;40.69;21.74;28;19.31
2016 - Polycube simplification for coarse layouts of surfaces and volumes_femur_polycube_in;yes;145;36;63;59;40.69;21.74;28;19.31
2016 - Structured Volume Decomposition via Generalized Sweeping_rockerarm_3;yes;82;16;57;53;64.63;22.83;16;19.51
2012 - All-Hex Meshing using Singularity-Restricted Field_sculpture-A;yes;51;11;27;21;41.18;12.86;10;19.61
2011 - All-Hex Mesh Generation via Volumetric PolyCube Deformation_asm001;yes;122;28;68;64;52.46;14.29;24;19.67
2014 - l1-Based Construction of Polycube Maps from Complex Shapes_rod;yes;122;30;70;64;52.46;14.29;24;19.67
2017 - Explicit Cylindrical Maps for General Tubular Shapes_cactus;yes;81;21;57;51;62.96;21.54;16;19.75
2014 - l1-Based Construction of Polycube Maps from Complex Shapes_buste;yes;361;99;203;166;45.98;21.07;72;19.94
2015 - Practical Hex-Mesh Optimization via Edge-Cone Rectification_hanger_stresstest_in;yes;50;10;28;21;42;15.94;10;20
2015 - Practical Hex-Mesh Optimization via Edge-Cone Rectification_hanger_stresstest_out;yes;50;10;28;21;42;15.94;10;20
2016 - Polycube simplification for coarse layouts of surfaces and volumes_chinese_dragon_polycube_out;yes;295;70;151;132;44.75;18.73;59;20
2016 - Polycube simplification for coarse layouts of surfaces and volumes_teapot_model_out;yes;193;43;98;88;45.6;21.76;39;20.21
2016 - Polycube simplification for coarse layouts of surfaces and volumes_teapot_polycube_out;yes;193;43;98;88;45.6;21.76;39;20.21
2016 - Polycube simplification for coarse layouts of surfaces and volumes_block_model_in;yes;162;42;109;105;64.81;8.66;33;20.37
2016 - Polycube simplification for coarse layouts of surfaces and volumes_block_polycube_in;yes;162;42;109;105;64.81;8.66;33;20.37
2019 - Selective Padding for Polycube‐Based Hexahedral Meshing_Lego_L0;yes;983;236;627;604;61.44;13.82;205;20.85
2019 - Singularity Structure Simplification of Hexahedral Mesh via Weighted Ranking_eight_input;yes;3867;1076;2250;1963;50.76;24.04;813;21.02
2016 - Structured Volume Decomposition via Generalized Sweeping_twistedu;yes;19;4;19;14;73.68;23.29;4;21.05
2016 - Skeleton-driven Adaptive Hexahedral Meshing of Tubular Shapes_armadillo;yes;301;82;180;165;54.82;7.49;64;21.26
2014 - l1-Based Construction of Polycube Maps from Complex Shapes_bunny;yes;273;73;157;129;47.25;21.89;59;21.61
2015 - Practical Hex-Mesh Optimization via Edge-Cone Rectification_bunny_in;yes;273;74;153;128;46.89;20.42;59;21.61
2015 - Practical Hex-Mesh Optimization via Edge-Cone Rectification_bunny_out;yes;273;74;153;128;46.89;20.42;59;21.61
2012 - All-Hex Meshing using Singularity-Restricted Field_bone;yes;87;15;50;43;49.43;29.27;19;21.84
2019 - Singularity Structure Simplification of Hexahedral Mesh via Weighted Ranking_fertility_output;yes;310;99;222;220;70.97;6.93;68;21.94
2016 - Polycube simplification for coarse layouts of surfaces and volumes_table2_polycube_in;yes;90;26;61;56;62.22;11.51;20;22.22
\end{filecontents*}
\csvreader[respect all, separator=semicolon, before reading=\begin{adjustbox}{max width=\linewidth}, after reading=\end{adjustbox},tabular=lrrrrrrrr, head to column names=true,
table head =
\bfseries Model & \bfseries BC &  \textcolor{gray}{BC\textsuperscript{--}} & \;\;\;\;\;\;\;\;\;\textcolor{gray}{raw} & \bfseries \!MC\textsuperscript{+}\!\! & \;\;\;\;\;\;\bfseries $\frac{\text{MC\textsuperscript{+}}}{\text{BC}}$ & \textcolor{gray}{\;\;\;\;\;\;\;\;\;\;\;\;\;T\;\;} & \;\;\;\;\;\;\bfseries MC & \!\!\bfseries $\frac{\text{MC}}{\text{BC}}$ \vspace{1mm}\\,
table foot =,
]{tables/StatsMay18part2.csv}{}
{{\textsc{\Model}}\;\;\;\;\;\;\;\; & \BC & \textcolor{gray}{\BCR} & \textcolor{gray}{\MCraw} & \MCplus & \!\MCpluspercent\% & \textcolor{gray}{\Tpercent\%} & \MC & \!\MCpercent\%}
\vspace{-0.1cm}
\caption{Statistics on a dataset of
{hexahedral meshes}. Numbers of blocks in the base complex (BC), reduced base complex (BC\textsuperscript{--}), raw motorcycle complex (raw), reduced motorcycle complex with preserved singularity-adjacent walls (MC\textsuperscript{+}), fully reduced motorcycle complex (MC), percentage of arcs that are T-arcs (T).\label{fig:stats}}
\vspace{-0.45cm}
\end{table*}

\addtocounter{table}{-1}    

\begin{table*}[h!]
\rowcolors{1}{}{backgray}
\begin{filecontents*}{tables/StatsMay18part3.csv}
Model;hexMeshValid;BC;BCR;MCraw;MCplus;MCpluspercent;Tpercent;MC;MCpercent
2019 - Selective Padding for Polycube‐Based Hexahedral Meshing_Chamfer_L4;yes;22;5;11;11;50;6.02;5;22.73
2014 - l1-Based Construction of Polycube Maps from Complex Shapes_kitty;yes;121;28;68;59;48.76;23.13;28;23.14
2016 - Structured Volume Decomposition via Generalized Sweeping_rockerarm_2;yes;82;17;57;53;64.63;22.83;19;23.17
2019 - Singularity Structure Simplification of Hexahedral Mesh via Weighted Ranking_eight_output;yes;43;6;35;35;81.4;15;10;23.26
2016 - Structured Volume Decomposition via Generalized Sweeping_Kitten_with_bifur_padded;yes;82;16;57;53;64.63;22.83;20;24.39
2016 - Structured Volume Decomposition via Generalized Sweeping_rockerarm_1;yes;82;18;57;53;64.63;22.83;20;24.39
2016 - Structured Volume Decomposition via Generalized Sweeping_hand7_padded;yes;159;36;102;91;57.23;23.29;39;24.53
2019 - Selective Padding for Polycube‐Based Hexahedral Meshing_Wrench;yes;61;13;32;30;49.18;6.48;15;24.59
2012 - All-Hex Meshing using Singularity-Restricted Field_sculpture-B;yes;69;16;33;33;47.83;10.17;17;24.64
2019 - Singularity Structure Simplification of Hexahedral Mesh via Weighted Ranking_toy1_output;yes;144;35;113;104;72.22;11.09;36;25
2019 - Dual Sheet Meshing - An Interactive Approach to Robust Hexahedralization_bunny;yes;34;7;20;17;50;8.89;9;26.47
2016 - Skeleton-driven Adaptive Hexahedral Meshing of Tubular Shapes_dinopet;yes;117;31;113;109;93.16;5.98;31;26.5
2016 - Skeleton-driven Adaptive Hexahedral Meshing of Tubular Shapes_big_buddy;yes;124;36;101;88;70.97;10.89;33;26.61
2016 - Skeleton-driven Adaptive Hexahedral Meshing of Tubular Shapes_rocker_arm;yes;45;13;39;39;86.67;10.34;12;26.67
2019 - Selective Padding for Polycube‐Based Hexahedral Meshing_Chamfer_L0;yes;29;8;18;18;62.07;2.48;8;27.59
2016 - All-Hex Meshing Using Closed-Form Induced Polycube_knot-hex;yes;50;16;40;40;80;15.79;14;28
2014 - l1-Based Construction of Polycube Maps from Complex Shapes_angel_3;yes;78;22;64;54;69.23;17.8;22;28.21
2012 - All-Hex Meshing using Singularity-Restricted Field_ellipsoid-B;yes;7;2;7;7;100;0;2;28.57
2012 - All-Hex Meshing using Singularity-Restricted Field_ellipsoid-C;yes;7;2;7;7;100;0;2;28.57
2016 - Skeleton-driven Adaptive Hexahedral Meshing of Tubular Shapes_warrior;yes;287;77;262;254;88.5;5.99;82;28.57
2016 - Skeleton-driven Adaptive Hexahedral Meshing of Tubular Shapes_blood_vessel;yes;52;17;48;48;92.31;5.77;15;28.85
2017 - Hexahedral Mesh Generation via Constrained Quadrilateralization_cube;yes;17;5;16;13;76.47;8.7;5;29.41
2017 - Hexahedral Mesh Generation via Constrained Quadrilateralization_pone.0177603.s001;yes;17;5;14;13;76.47;8.7;5;29.41
2012 - All-Hex Meshing using Singularity-Restricted Field_double;yes;71;19;52;41;57.75;25.25;21;29.58
2016 - Skeleton-driven Adaptive Hexahedral Meshing of Tubular Shapes_cactus;yes;37;11;33;33;89.19;8.11;11;29.73
2016 - Skeleton-driven Adaptive Hexahedral Meshing of Tubular Shapes_clef;yes;37;11;33;33;89.19;8.11;11;29.73
2016 - Skeleton-driven Adaptive Hexahedral Meshing of Tubular Shapes_dino;yes;47;14;47;47;100;0;14;29.79
2016 - Skeleton-driven Adaptive Hexahedral Meshing of Tubular Shapes_octopus;yes;104;30;75;69;66.35;9.68;31;29.81
2016 - Structured Volume Decomposition via Generalized Sweeping_hand7_3;yes;67;20;46;40;59.7;16.38;20;29.85
2016 - Polycube simplification for coarse layouts of surfaces and volumes_hand_model_out;yes;107;32;74;70;65.42;14.33;32;29.91
2016 - Polycube simplification for coarse layouts of surfaces and volumes_hand_polycube_out;yes;107;32;74;70;65.42;14.33;32;29.91
2017 - Explicit Cylindrical Maps for General Tubular Shapes_femur_shell1;yes;30;8;24;22;73.33;11.76;9;30
2017 - Explicit Cylindrical Maps for General Tubular Shapes_femur_shell2;yes;30;8;24;22;73.33;11.76;9;30
2019 - Singularity Structure Simplification of Hexahedral Mesh via Weighted Ranking_toy2_output;yes;129;34;94;88;68.22;12.47;39;30.23
2019 - Selective Padding for Polycube‐Based Hexahedral Meshing_Gear;yes;85;24;46;45;52.94;7.06;26;30.59
2016 - Polycube simplification for coarse layouts of surfaces and volumes_cubespikes_model_out;yes;111;28;107;101;90.99;9.35;34;30.63
2016 - Polycube simplification for coarse layouts of surfaces and volumes_cubespikes_polycube_out;yes;111;28;107;101;90.99;9.35;34;30.63
2015 - Practical Hex-Mesh Optimization via Edge-Cone Rectification_block_in;yes;100;31;100;98;98;2.08;31;31
2015 - Practical Hex-Mesh Optimization via Edge-Cone Rectification_block_out;yes;100;31;100;98;98;2.08;31;31
2015 - Practical Hex-Mesh Optimization via Edge-Cone Rectification_block_stresstest_in;yes;100;31;100;98;98;2.08;31;31
2015 - Practical Hex-Mesh Optimization via Edge-Cone Rectification_block_stresstest_out;yes;100;31;100;98;98;2.08;31;31
2016 - Structured Volume Decomposition via Generalized Sweeping_hand7_1;yes;67;21;46;40;59.7;17.8;21;31.34
2016 - Structured Volume Decomposition via Generalized Sweeping_hand7_2;yes;67;19;46;40;59.7;17.8;21;31.34
2016 - Polycube simplification for coarse layouts of surfaces and volumes_block_model_out;yes;100;33;100;98;98;2.08;33;33
2016 - Polycube simplification for coarse layouts of surfaces and volumes_block_polycube_out;yes;100;33;100;98;98;2.08;33;33
2019 - Symmetric Moving Frames_hex_brokenbullet;yes;24;7;20;17;70.83;12.37;8;33.33
2016 - Structured Volume Decomposition via Generalized Sweeping_kitten_with_bifur_1;yes;26;5;22;22;84.62;10.53;9;34.62
2019 - Dual Sheet Meshing - An Interactive Approach to Robust Hexahedralization_rod;yes;43;16;40;36;83.72;8.18;15;34.88
2016 - Efficient Volumetric PolyCube-Map Construction_Sphinx_hex-f;yes;30;10;23;23;76.67;11.11;11;36.67
2016 - Skeleton-driven Adaptive Hexahedral Meshing of Tubular Shapes_fertility;yes;49;18;45;41;83.67;15.56;18;36.73
2017 - Explicit Cylindrical Maps for General Tubular Shapes_femur;yes;38;12;32;30;78.95;16.44;14;36.84
2016 - Skeleton-driven Adaptive Hexahedral Meshing of Tubular Shapes_santa;yes;42;13;40;38;90.48;8.24;16;38.1
2016 - Skeleton-driven Adaptive Hexahedral Meshing of Tubular Shapes_block;yes;36;14;36;34;94.44;5.88;14;38.89
2016 - Polycube simplification for coarse layouts of surfaces and volumes_table2_polycube_out;yes;59;19;58;53;89.83;8.62;23;38.98
2016 - Structured Volume Decomposition via Generalized Sweeping_cat_2;yes;5;2;5;5;100;0;2;40
2016 - Structured Volume Decomposition via Generalized Sweeping_cat_3;yes;5;2;5;5;100;0;2;40
2016 - Structured Volume Decomposition via Generalized Sweeping_Dolphin_1;yes;5;2;5;5;100;0;2;40
2016 - Structured Volume Decomposition via Generalized Sweeping_Dolphin_2;yes;5;2;5;5;100;0;2;40
2016 - Structured Volume Decomposition via Generalized Sweeping_Dolphin_3;yes;5;2;5;5;100;0;2;40
2016 - Structured Volume Decomposition via Generalized Sweeping_femur1_2;yes;5;2;5;5;100;0;2;40
2016 - Structured Volume Decomposition via Generalized Sweeping_femur1_3;yes;5;2;5;5;100;0;2;40
2016 - Structured Volume Decomposition via Generalized Sweeping_kitten_2;yes;5;2;5;5;100;40;2;40
2016 - Structured Volume Decomposition via Generalized Sweeping_rabbit_2;yes;5;2;5;5;100;0;2;40
2016 - Structured Volume Decomposition via Generalized Sweeping_rabbit_3;yes;5;2;5;5;100;0;2;40
2017 - Explicit Cylindrical Maps for General Tubular Shapes_cylinder_mixed;yes;7;3;7;7;100;0;3;42.86
2017 - Explicit Cylindrical Maps for General Tubular Shapes_cylinder_mixed_w_torsion;yes;7;3;7;7;100;0;3;42.86
2019 - Dual Sheet Meshing - An Interactive Approach to Robust Hexahedralization_bone;yes;16;6;12;12;75;4.26;7;43.75
2016 - Structured Volume Decomposition via Generalized Sweeping_bunny_1;yes;18;8;16;16;88.89;6.74;8;44.44
2016 - Structured Volume Decomposition via Generalized Sweeping_bunny_2;yes;18;8;16;16;88.89;6.74;8;44.44
2016 - Structured Volume Decomposition via Generalized Sweeping_bunny6_1;yes;18;8;16;16;88.89;6.74;8;44.44
2016 - Structured Volume Decomposition via Generalized Sweeping_bunny6_2;yes;18;8;16;16;88.89;6.74;8;44.44
2016 - Structured Volume Decomposition via Generalized Sweeping_bunny6_3;yes;18;8;16;16;88.89;6.74;8;44.44
2019 - Selective Padding for Polycube‐Based Hexahedral Meshing_Column;yes;18;8;18;17;94.44;3.85;8;44.44
2016 - All-Hex Meshing Using Closed-Form Induced Polycube_nut-hex;yes;12;6;12;12;100;0;6;50
2019 - Symmetric Moving Frames_hex_tetrahedron;yes;4;2;4;4;100;0;2;50
2017 - Explicit Cylindrical Maps for General Tubular Shapes_spring;yes;7;4;7;7;100;0;4;57.14
2016 - All-Hex Meshing Using Closed-Form Induced Polycube_fancy_ring-hex;yes;5;3;5;5;100;14.29;3;60
2016 - Structured Volume Decomposition via Generalized Sweeping_cat_1;yes;5;3;5;5;100;0;3;60
2016 - Structured Volume Decomposition via Generalized Sweeping_femur1_1;yes;5;3;5;5;100;0;3;60
2016 - Structured Volume Decomposition via Generalized Sweeping_kitten_1;yes;5;3;5;5;100;40;3;60
2016 - Structured Volume Decomposition via Generalized Sweeping_rabbit_1;yes;5;3;5;5;100;0;3;60
2019 - Dual Sheet Meshing - An Interactive Approach to Robust Hexahedralization_double-torus;yes;5;3;5;5;100;0;3;60
2017 - Explicit Cylindrical Maps for General Tubular Shapes_femur_shell0;yes;8;5;8;8;100;0;5;62.5
2016 - All-Hex Meshing Using Closed-Form Induced Polycube_KPDloekr-hex;yes;2;2;2;2;100;38.46;2;100
2017 - Explicit Cylindrical Maps for General Tubular Shapes_cylinder_grid;yes;1;1;1;1;100;0;1;100
2017 - Explicit Cylindrical Maps for General Tubular Shapes_cylinder_polar;yes;1;1;1;1;100;0;1;100
2018 - Fuzzy clustering based pseudo-swept volume decomposition for hexahedral meshing_Example_5;yes;1;1;1;1;100;0;1;100
\end{filecontents*}
\csvreader[respect all, separator=semicolon, before reading=\begin{adjustbox}{max width=\linewidth}, after reading=\end{adjustbox},tabular=lrrrrrrrr, head to column names=true,
table head =
\bfseries Model & \bfseries BC &  \textcolor{gray}{BC\textsuperscript{--}} & \;\;\;\;\;\;\;\;\;\textcolor{gray}{raw} & \bfseries \!MC\textsuperscript{+}\!\! & \;\;\;\;\;\;\bfseries $\frac{\text{MC\textsuperscript{+}}}{\text{BC}}$ & \textcolor{gray}{\;\;\;\;\;\;\;\;\;\;\;\;\;T\;\;} & \;\;\;\;\;\;\bfseries MC & \!\!\bfseries $\frac{\text{MC}}{\text{BC}}$ \vspace{1mm}\\,
table foot =,
]{tables/StatsMay18part3.csv}{}
{{\textsc{\Model}}\;\;\;\;\;\;\;\;\;\;\;\;\;\;\;\;\;\;\;\;\;\;\;\; & \BC & \textcolor{gray}{\BCR} & \textcolor{gray}{\MCraw} & \MCplus & \!\MCpluspercent\% & \textcolor{gray}{\Tpercent\%} & \MC & \!\MCpercent\%}
\vspace{-0.1cm}
\caption{Statistics on a dataset of
{hexahedral meshes}. Numbers of blocks in the base complex (BC), reduced base complex (BC\textsuperscript{--}), raw motorcycle complex (raw), reduced motorcycle complex with preserved singularity-adjacent walls (MC\textsuperscript{+}), fully reduced motorcycle complex (MC), percentage of arcs that are T-arcs (T).\label{fig:stats}}
\vspace{-0.45cm}
\end{table*}

\begin{figure*}[tbh]
\centering
  \includegraphics[width=0.92\textwidth]{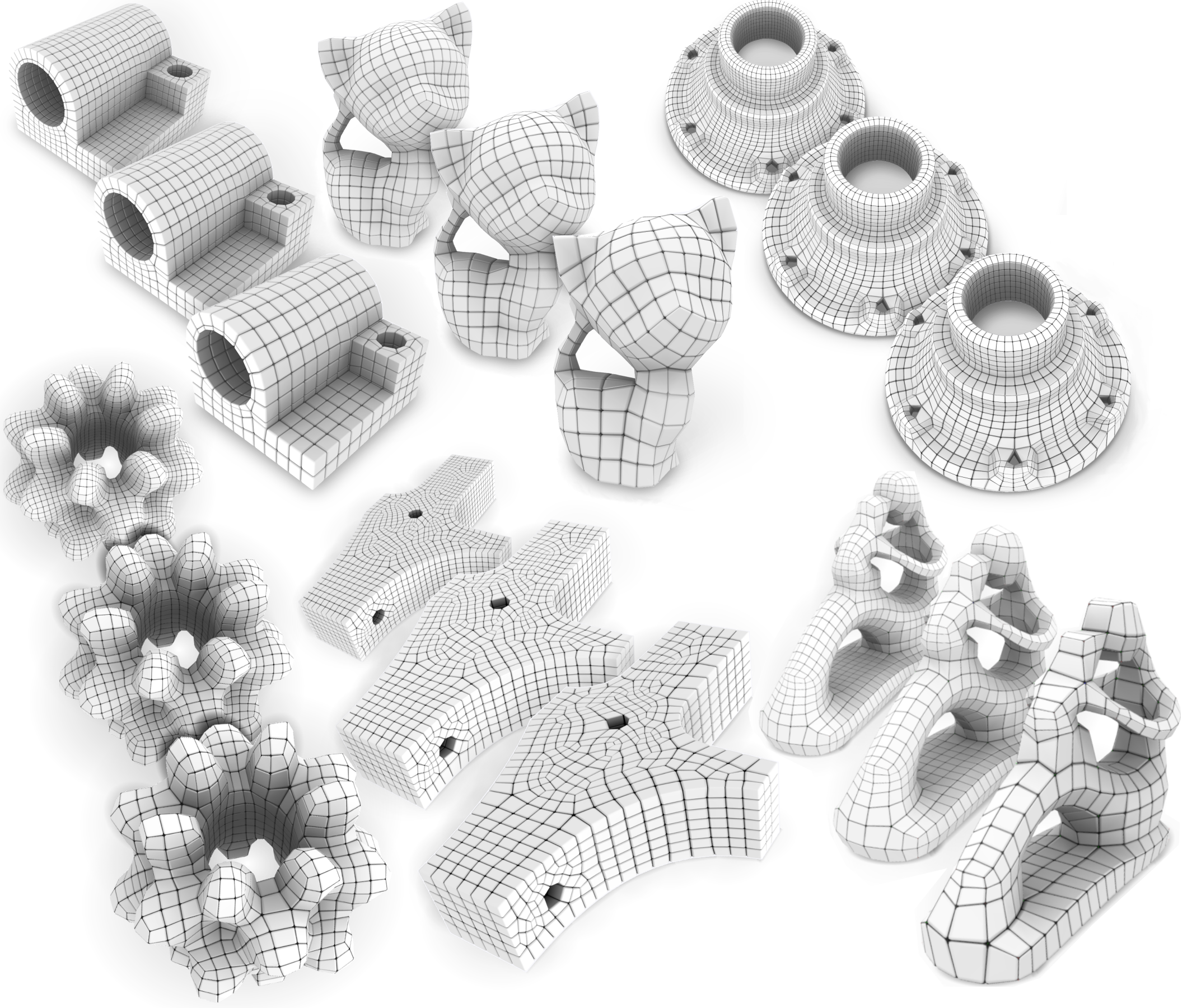}
    \vspace{-0.15cm}
  \caption{Hexahedral meshes of varying density generated by the described quantization method based on the proposed motorcycle complex.
  \label{fig:quantization}}
  \vspace{-0.1cm}
\end{figure*}

\section{Quantization Control}

In addition to Fig.~15, Fig.~\ref{fig:quantization} shows further examples of the ability to control the density of hexahedral meshes resulting from our quantization procedure in a fine-grained manner. To extend the range of test cases beyond those available in the dataset from Table 2 for these examples, we reconstructed seamless parametrizations from given hexahedral meshes: a tetrahedral mesh was generated, containing the singular edges, and a parametrization was imposed on it under which each hex is a unit cube. These seamlessly parametrized tetrahedral meshes can then be taken as input to the quantization procedure like those from the Table 2 dataset.

\begin{figure}[b]
  \vspace{-0.25cm}
 \begin{tikzpicture}
 \begin{axis}[
     axis background/.style={fill=gray!7},
     axis line style={white},
     every x tick/.style={white},
     every y tick/.style={white},
     legend style={draw=none},
     legend style={at={(0.075,0.94)},anchor=north west},
     reverse legend,
     legend cell align={left},
     yticklabel = \pgfmathprintnumber{\tick}\,K,
     grid=both,
     major grid style={line width=2pt,draw=white},
     height=47mm,
     width=87mm,
     xlabel={quantization factor ($s^3$)},
     ylabel={\!number of hexes ($n$)},
     xmin=0, xmax=25,
     ymin=0, ymax=124,
     xtick={0,5,10,15,20,25},
     ytick={0,20,40,60,80,100,120},
     xlabel style={yshift=1mm},
     ylabel style={yshift=-2mm},
     ticklabel style = {font=\small},
 ]
 \addplot+[plotyellow, very thick, no marks] table[y expr=\thisrowno{1}/1000]{images/plot/example_2_output_x3.txt};
 \addplot+[const plot, plotgreen, very thick, no marks] table[y expr=\thisrowno{1}/1000]{images/plot/Example_2_MCG_HB.txt};
 \addplot+[const plot, plotblue, very thick, no marks] table[y expr=\thisrowno{1}/1000]{images/plot/example_2_output_bc.txt};

 \addlegendentry{\smash{$n \sim s^3$}\phantom{C}}
 \addlegendentry{MC-based}
 \addlegendentry{BC-based}

 \end{axis}
 \end{tikzpicture}
 \vspace{-0.6cm}
 \caption{Mesh resolution can be controlled more finely when using the MC, not the BC, as basis for quantization using standard objective (10). This is illustrated here for model \textsc{Example\_2} (Fig.~\ref{fig:quantization} bottom center). 
 The number of hexahedra in the extracted hex mesh is shown versus the scaling factor (i.e., target hex edge length is $\nicefrac{1}{s}$). The gray line is the conceptual optimum: the number of resulting hexes exactly antiproportional to the target hex volume.
 }\label{fig:plot}
 \end{figure}
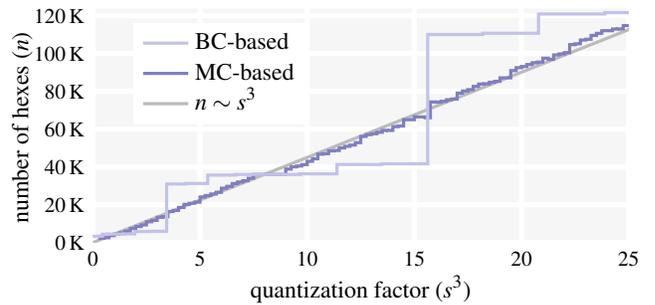

Fig.~\ref{fig:plot} illustrates that when basing the quantization system on the base complex rather than the motorcycle complex, density control may be significantly less fine-grained. This is intimately related to the lower number of degrees of freedom in the conforming structure of the base complex (cf.~Fig. 14).

\clearpage
\newpage

%% file: text/00frontmatter.tex
\teaser{
\vspace{-0.15cm}

\begin{overpic}[width=.97\textwidth]{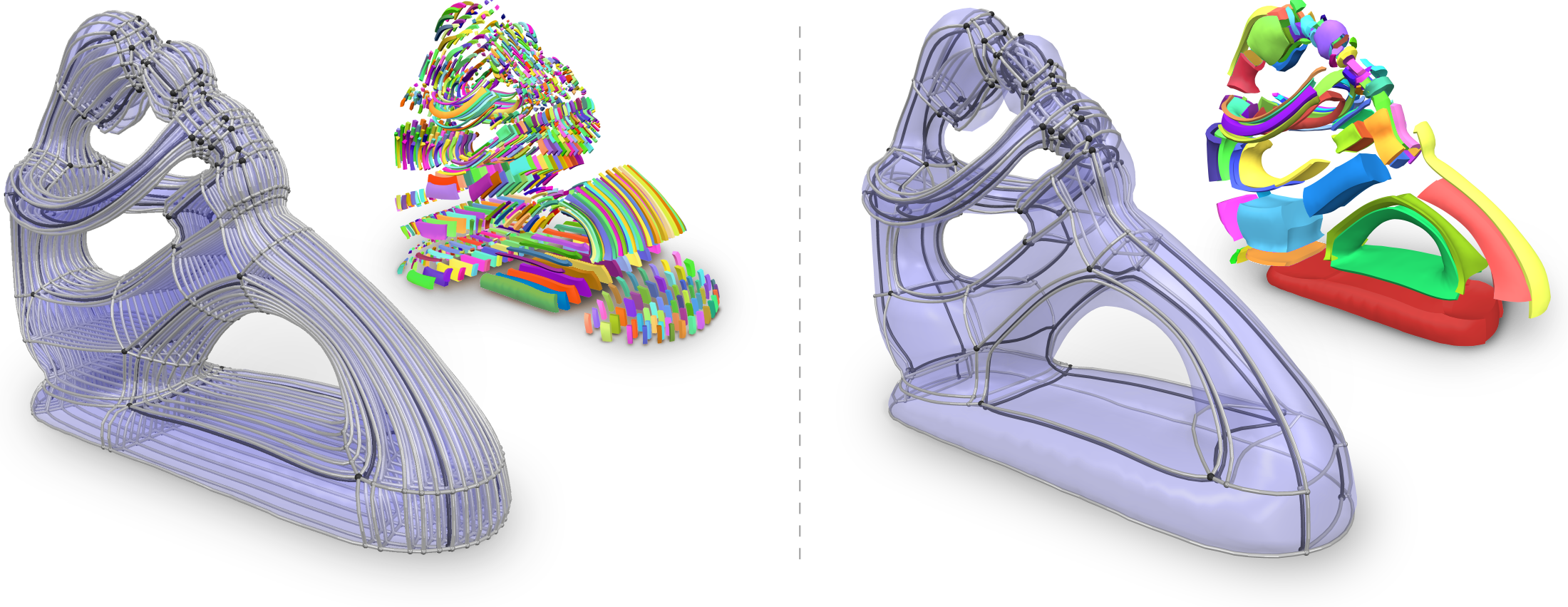}
\put(33.5,13.0){\small \textbf{Base Complex}}
\put(33.5,11.0){\small 1352 blocks}
\put(87.65,13.0){\small \textbf{Motorcycle Complex}}
\put(87.65,11.0){\small 79 blocks \;(5.8\%)}
\end{overpic}
\vspace{-0.35cm}
\caption{Base Complex (left) and our Motorcycle Complex (right) induced by the same volumetric seamless parametrization of a solid object, both providing a structured partition into cuboid blocks. The motorcycle complex often partitions the object's interior into a much smaller number of blocks, here just 5.8\%, 79 instead of 1352 blocks (see the exploded views). We provide a definition of this motorcycle complex, describe algorithms for its construction, and demonstrate its use and benefits.\label{fig:teaser}}
}

\maketitle

\begin{abstract}
The so-called motorcycle graph has been employed in recent years for various purposes in the context of structured and aligned block decomposition of 2D shapes and 2-manifold surfaces. Applications are in the fields of surface parametrization, spline space construction, semi-structured quad mesh generation, or geometry data compression.
We describe a generalization of this motorcycle graph concept to the three-dimensional volumetric setting. Through careful extensions aware of topological intricacies of this higher-dimensional setting, we are able to guarantee important block decomposition properties also in this case.
We describe algorithms for the construction of this 3D motorcycle complex on the basis of either hexahedral meshes or seamless volumetric parametrizations. Its utility is illustrated on examples in hexahedral mesh generation and volumetric T-spline construction.

%
%


\printccsdesc
\end{abstract}

%% file: text/01introduction.tex
\section{Introduction}

The motorcycle graph \cite{eppstein2008motorcycle,eppstein1999raising} has been used in various computer graphics and geometry processing applications to partition surfaces in a structured manner, as discussed further in Sec.~\ref{sec:related}. 
Conceptually, a number of particles (called \emph{motorcycles}) are traced over a surface, each one stopping when reaching a trace. The collection of traces finally forms a surface-embedded graph that partitions the surface.
This idea has been used on surfaces equipped with various structures that define the directions the motorcycles take, most relevantly:
\begin{itemize}
\item cross fields or frame fields,
\item seamless or integer-grid parametrizations,
\item quadrilateral meshes.
\end{itemize}
These objects all impose a 
structure on the surface that defines four directions everywhere,
except at a number of isolated singularities. Under mild assumptions, the motorcycle graph, with particles starting at these singularities, yields a partition of the surface into patches that all are \emph{four-sided}, completely \emph{regular} in their interior, and \emph{aligned} with the field's streamlines, the parametrization's isolines, or the mesh's edges, respectively.

\begin{figure}[t]
\centering
\begin{overpic}[width=0.98\linewidth]{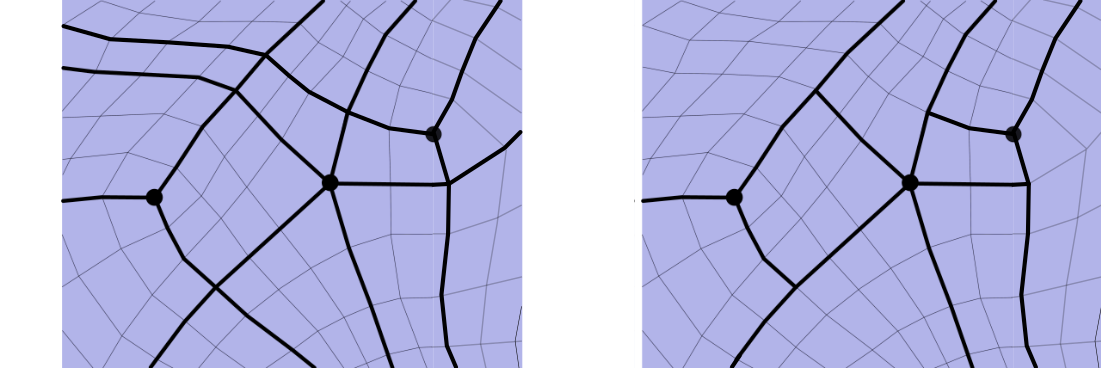}
\put(0.55,5.5){\begin{turn}{90}{Base Complex}\end{turn}}
\put(53.5,2.2){\begin{turn}{90}{Motorcycle Graph}\end{turn}}
\end{overpic}
\vspace{-0.15cm}
\caption{Left: base complex (black) of a quad mesh (grey edges), providing a conforming partition. Right: motorcycle graph,
providing a non-conforming partition, with 4~T-joints in this example.\label{fig:Tjoint}}
\vspace{-0.2cm}
\end{figure}

A related structure is the so-called base complex \cite{bommes2011global}, 
also referred to as quad layout \cite{campen2014quad,Pietroni:2016}. It can be obtained by not letting particles stop at traces (but only at singularities or boundaries). The base complex is known to be the coarsest \emph{conforming} partition into four-sided, regular, aligned patches.
By contrast, the partition obtained by the motorcycle graph is typically \emph{non-conforming}: there can be T-joints, as illustrated in Fig.~\ref{fig:Tjoint}.

For use cases that are able to handle this non-conformity (or even benefit from it), the motorcycle graph provides a major advantage: it is often much simpler (having a smaller number of patches), in some cases even by orders of magnitude, than the base complex. It has been exploited in recent years (see Sec.~\ref{sec:related}) for scenarios like
\begin{itemize}
\item generation of quad meshes \cite{Myles:2014:RFG},
\item localized structured remeshing \cite{nuvoli2019quadmixer},
\item quantization of global parametrizations \cite{Campen:2015:QGP},
\item construction of T-spline spaces \cite{campen2017similarity},
\item texture mapping of surfaces \cite{schertler2018generalized},
\item mesh-based computational fabrication \cite{ChebychevMCG}.
\end{itemize}

\vspace{-0.2cm}
\subsection{Contribution}

We propose the \emph{motorcycle complex}, a generalization of the 2D motorcycle graph to the 3D volumetric setting. In analogy to the 2D case, it partitions a volumetric object, equipped with a suitable directional structure, into regular cuboid blocks (rather than quadrilateral patches) in a non-conforming manner, cf.~Fig.~\ref{fig:teaser}.
Suitable guiding structures are volumetric seamless parametrizations, 3D integer-grid maps, and hexahedral meshes.

Algorithmic tools necessary to compute this motorcycle complex, based either on parametrized tetrahedral meshes or on hexahedral meshes, are introduced.
We analyze the characteristics of the resulting complex, and show that important properties are guaranteed by the induced partition. Most importantly, this includes the cuboidal structure and the regularity of each induced cell.

Looking at the ways the motorcycle graph has been successfully leveraged in the 2D case (in particular when it comes to guaranteeing robustness), this motorcycle complex has the potential to serve as foundation for important advances in hexahedral mesh generation, volumetric T-spline definition, and further problems related to grid generation, volumetric parametrization, and isogeometric analysis. We illustrate this with two example applications in Sec.~\ref{sec:results}.

\subsection{Overview}

\paragraph*{Input} 
Following Fig.~\ref{fig:overview}, the input to our method is a seamless parametrization (Sec.~\ref{sec:seamless}) on a tetrahedral mesh, sanitized numerically (Sec.~\ref{sec:sanity}) for robustness if necessary.
Alternatively a (non-parametrized) hexahedral mesh can be considered---analogously to how the motorcycle graph in 2D has been used on suitably parametrized triangle meshes as well as on quadrilateral meshes.
While the case of seamless parametrization input is most relevant (and algorithmically more challenging and interesting), we discuss the simpler hexahedral mesh case as a more intuitive entry, too.

\paragraph*{Goal} 
The input object is to be partitioned into a small number of cuboid blocks (see Fig.~\ref{fig:teaser}) such that they are regular in their interior (not containing any singularities or irregular vertices/edges, respectively) and each of a block's six boundary patches is aligned with an iso-surface in the parametrization or a sheet of quads in the hexahedral mesh, respectively. In other words, the blocks are axis-aligned rectangular cuboids under the parametrization, or regular
$l\!\times\! m \!\times\! n$-grid pieces of the hexahedral mesh, respectively.

\paragraph*{Approach} 
The goal is met by constructing a motorcycle complex induced by the input, as defined in Sec.~\ref{sec:themc}. This is done in three algorithmic steps (see~Fig.~\ref{fig:overview} right), detailed in Sec.~\ref{sec:impl}. In step 1, parts of the iso-surfaces incident at the singularities are incrementally designated as block walls in an expansion process. In step 2, possibly additional walls are designated to guarantee the desired block decomposition property. In step 3, redundant walls are retracted, in regions where the initial expansion process was over-zealous.

\begin{figure}[t]
\centering
\begin{overpic}[width=0.99\linewidth]{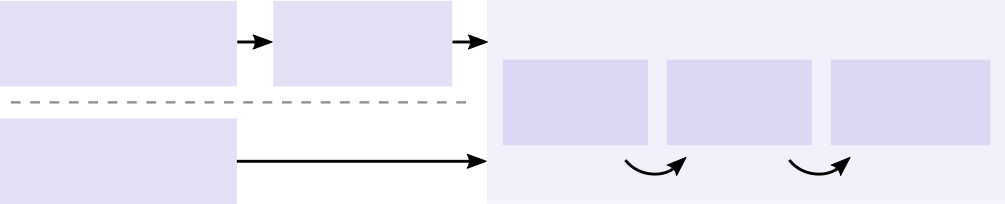}
\put(1.1,17.2){\small seamless}
\put(1.1,13.4){\small parametrization}

\put(1.1,5.2){\small hexahedral}
\put(1.1,1.4){\small mesh}

\put(28.25,17.2){\small numerical}
\put(28.25,13.4){\small sanitization}

\put(51,11){\small wall}
\put(51,7.5){\small tracing}

\put(67.5,11){\small toroid}
\put(67.5,7.5){\small splitting}

\put(83.7,11){\small wall}
\put(83.7,7.5){\small retraction}

\put(51.5,16.8){\small motorcycle complex construction}

\end{overpic}
\vspace{-0.1cm}
\caption{We take as input either a volumetric seamless parametri-zation (on a tetrahedral mesh) or a hexahedral mesh, and compute an induced motorcycle complex in three algorithmic steps.\label{fig:overview}}
\vspace{-0.25cm}
\end{figure}

%% file: text/02relatedwork.tex
\section{Related Work}
\label{sec:related}

The original 2D Euclidean definition of the motorcycle graph goes back to work by Eppstein and Erickson \cite{eppstein1999raising}. It was extended to curved surfaces, i.e., Riemannian manifolds, initially for the purpose of quadrilateral mesh partitioning \cite{eppstein2008motorcycle}. In this setting the mesh's edges provide the directional information guiding the motorcycles across the surfaces. This quad mesh driven motorcycle graph has been employed in further contexts, for
reverse engineering \cite{gunpinar2014motorcycle}, texture mapping \cite{schertler2018generalized}, computational fabrication \cite{ChebychevMCG}, and quadrilateral remeshing \cite{nuvoli2019quadmixer,razafindrazaka2017optimal}.

The idea of the motorcycle graph has been adapted to surfaces equipped with other directional guiding structures
besides a quadrilateral mesh structure.
In particular, motorcycles following streamlines of a cross field \cite{Vaxman:FieldsSTAR} have been used for the reliable generation of global seamless surface parametrizations \cite{Myles:2014:RFG}. Motorcycles following the isolines of such seamless parametrizations \cite{Kaelberer:QuadCover,Bommes:MIQ,Myles:2012}, in turn, have been used for the purpose of robust parametrization quantization \cite{Campen:2015:QGP,Lyon:2019:PQF,Lyon:2021,lyon2021simplerlayouts}. This results in integer-grid maps, which are important ingredients for the generation of quadrilateral meshes \cite{Bommes:RMIQ}. A generalized class of parametrizations, so-called seamless similarity parametrizations, provide another structure that can be used to guide motorcycles \cite{campen2017similarity}. This has been leveraged for the reliable construction of T-meshes that can serve as domain for the definition of T-spline spaces  \cite{campen2017similarity,karvciauskas2017t}.

For the 3D volumetric case, a concept analogous to the 2D motorcycle graph has not been described yet.
Generalizations of the above mentioned guiding structures,
however, often do exist.
Hexahedral meshes 
can be considered the natural generalization of quadrilateral meshes to the next dimension. Cross fields generalize to octahedral fields \cite{solomon2017boundary,huang2011boundary,liu2018singularity,Corman:2019:SMF,Zhang2020Octa}, and seamless parametrizations of triangular surface meshes extend naturally to tetrahedral volume meshes as well \cite{Nieser:2011}, see also Sec.~\ref{sec:seamless}.

So far only the base complex \cite[§2.2]{bommes2011global} (which also provides an aligned decomposition into regular blocks) has been considered in a 3D setting, for the case of hexahedral meshes \cite{gao2015hexahedral}. For hexahedral meshes with many details, this structure can be highly complex; even more so for seamless volume parametrizations (which can be viewed as infinitely dense hexahedral meshes), where it can easily become impractically large.

More distantly related are volumetric block decomposition algorithms not driven by a prescribed singularity structure
or targeting other use cases, based on plastering \cite{staten2010unconstrained}, medial axes  
\cite{sheffer1999hexahedral},
or cut sheets \cite{DualSheets,LoopyCuts}.
\vspace{-0.1cm}

%% file: text/03background.tex
\section{Background}

\subsection{Seamless Parametrization}
\label{sec:seamless}

Given a surface $M$, a \emph{seamless (surface) parametrization} \cite{Myles:2012} is a chart-based map $\phi: M^c \rightarrow \mathbb{R}^2$ (where $M^c$ is $M$ cut to one or more topological disks) such that chart transitions are rigid, with a rotation by some multiple of $\nicefrac{\pi}{2}$.
Analogously, given a volume $M$, a \emph{seamless (volume) parametrization} is a chart-based map $\phi: M^c \rightarrow \mathbb{R}^3$ (where $M^c$ is $M$ cut to one or more topological balls) such that chart transitions are rigid, with a rotation from the octahedral rotation group \cite{Nieser:2011}.

In the discrete 3D case, with $M$ given as a tetrahedral mesh, we assume $\phi$ to be affine per tetrahedron, with transitions across facets. Unless stated otherwise, a seamless parametrization is assumed to be \emph{valid} (non-degenerate and orientation preserving) and such that boundary facets of $M$ are aligned. A facet (edge) is said to be \emph{aligned} if its image under $\phi$ is constant in one (two) coordinate components, i.e., it is parallel to one coordinate plane (axis).
The sum of incident parametric dihedral angles around an edge of $M$ is a multiple of $\nicefrac{\pi}{2}$; an edge is \emph{regular} if it is $2\pi$ for interior, or $\pi$ for boundary edges; otherwise it is \emph{singular}.
Like all known use cases we require singular edges to be aligned.
For the practically by far most relevant types of singularities, deviating from the regular case by $\pm\nicefrac{\pi}{2}$ \cite{liu2018singularity}, this is inherent anyway \cite{ebke2013qex}; for generic singular edges constraints can ensure it \cite{Nieser:2011}.

\begin{figure}[t]
\centering
\begin{overpic}[width=0.98\linewidth]{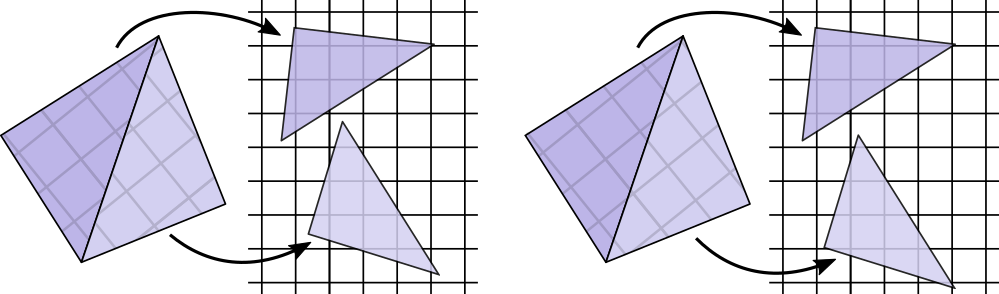}
\end{overpic}
\vspace{-0.05cm}
\caption{Illustration of a 2D seamless parametrization. Left: \emph{continuous}. Right: \emph{quantized}. In both cases, there is no scale discontinuity across the chart transition between the two triangles, and the rotation is some multiple of $\nicefrac{\pi}{2}$. On the right, additionally, the translation is integral, making the integer grid continuous.\label{fig:param}}
\vspace{-0.15cm}
\end{figure}

An \emph{integer-grid parametrization} \cite{Bommes:RMIQ} (also \emph{quantized} parametrization \cite{Campen:2015:QGP}) is a special case (Fig.~\ref{fig:param}): the translational components of all transitions as well as the constant components of all images of singular and aligned elements are from $\mathbb{Z}$ instead of $\mathbb{R}$.
Such an integer-grid parametrization (in the 3D case) naturally induces a hexahedral mesh \cite{Nieser:2011}. Singularities induce irregularities in the mesh (edges with more or less than 4 adjacent hexahedra), and sheets of quads in the mesh coincide with iso-surfaces of the parametrization.

We do not make any implicit assumption about parametrizations being quantized in the following. Where explicit distinction is necessary, we use the terms \emph{continuous} parametrization versus \emph{quantized} parametrization. Importantly, our method is able to operate on arbitrary continuous seamless parametrizations. As discussed in Sec.~\ref{sec:results} the motorcycle complex can actually be used to yield a quantized parametrization (and therefore also a hex mesh) from a continuous one---a task hard to solve with previous techniques.

\paragraph*{Metric}
We will argue about curves or surfaces in $M$ being straight, planar, parallel, or orthogonal \emph{with respect to} $\phi$ or \emph{in the $\phi$-metric}. This is to be understood as measuring these objects' images under $\phi$ in $\mathbb{R}^3$---or equivalently: measuring in $M$ using the metric tensor that is the pull-back through $\phi$ of the Euclidean metric tensor.

\begin{figure*}[t!]
  \includegraphics[width=.99\textwidth]{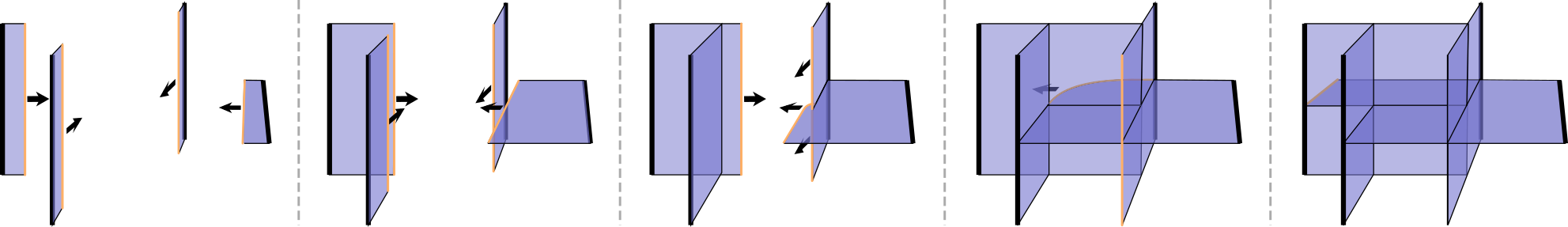}
  \vspace{-0.2cm}
  \caption{Illustration of the brush fire process in 3D. For simplicity and visual clarity only a single iso-surface per singular curve (bold black) is shown, clipped to a cubical region, in a setting where iso-surfaces are planar.
  The fire front is highlighted in orange, its conceptual direction of expansion is indicated by arrows. The supplementary video gives an animated impression of the process in more complex settings.\label{fig:snapshotsSymbolic}}
  \vspace{-0.25cm}
\end{figure*}

\subsection{2D Motorcycle Graph}
\label{sec:mc2d}

Various incarnations of the 2D motorcycle graph idea have been used on surfaces. 
For the case of a seamless parametrization $\phi$ providing guidance on surface $M$, it can be summarized as follows.
At each point $p_i \in M$ where $\phi$ is singular, for each direction $d_i$ of an incident iso-line of $\phi$ a particle $(p_i, d_i)$ is placed.
Simultaneously, each particle starts tracing (with unit speed, from $p_i$ in direction $d_i$) a curve across $M$ that is straight with respect to $\phi$, i.e., it is an iso-curve (taking transitions into account). A particle stops when it hits: (i) a trace (left behind by itself or another particle), (ii) a point where $\phi$ is singular, or (iii) the boundary of $M$.
Upon termination, the collection of traces forms a surface-embedded graph, the motorcycle graph. When instead 
{ignoring stopping criterion (i),} the resulting graph is the base complex (see Fig.~\ref{fig:Tjoint}). Clearly, the motorcycle graph is a subgraph of the base complex graph.

\begin{figure}[b]
\vspace{-0.1cm}
\includegraphics[width=.95\columnwidth]{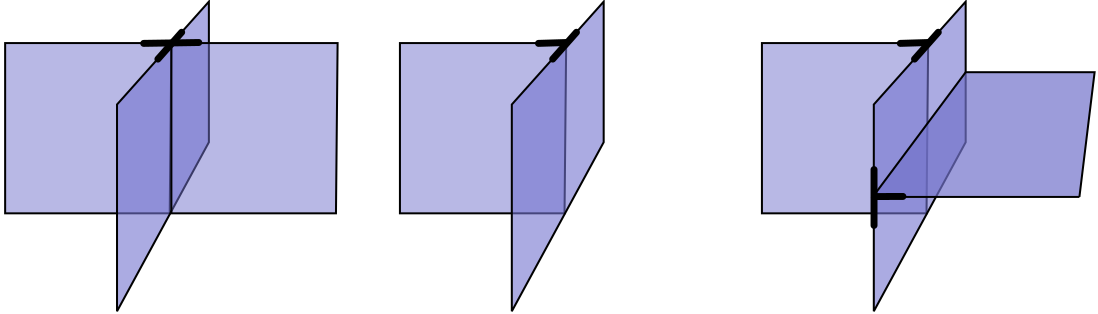}
\vspace{-0.3cm}
\caption{Left: regular crossing, formed by four walls. Center: T-joint, formed by three walls. Right: interaction of two T-joints; in contrast to the other two configurations (essentially extrusions of their 2D counterparts) this is a configuration specific to the 3D case.\label{fig:Ts}}
\end{figure}

The following properties were shown for the (non-empty) motorcycle graph \cite{eppstein2008motorcycle}:
\begin{itemize}
\item each patch has disk topology,
\item each patch has four sides, aligned with isolines,
\item each patch is regular, i.e., free of interior singularities.
\end{itemize}
Furthermore, the number of patches is within a constant factor of the minimum number possible for any partition with these properties.
Finding a truly minimal partition is known to be much harder than computing the motorcycle graph \cite{eppstein2008motorcycle}.

\vspace{-0.15cm}

%% file: text/04motorcyclecomplex.tex
\section{The Motorcycle Complex}
\label{sec:themc}

The idea behind the motorcycle graph does not generalize easily to higher dimensions.
While in 2D \emph{curves} (traces of particles) are sufficient to partition the two-manifold into patches, in 3D \emph{surfaces} are required to partition the manifold into blocks. These cannot be modeled as traces of some finite number of moving point particles. 
Instead, we interpret the construction process as an equivalent brush fire expansion process, in such a way that it is dimension-generic. Conceptually, a fire is ignited simultaneously at all points on singularities
of $\phi$. It is confined to spread within $(n\!-\!1)$-dimensional isoparametric submanifolds that contain the singularities, and cannot cross points already burnt. If $M$ has a boundary,
it is additionally considered burnt.

For $n=2$ the singularities are points and the $1$-dimensional isoparametric submanifolds are iso-curves of $\phi$, i.e.,  curves $c(t)$ along which $\phi(c(t)) = (u,v)$ is constant in either $u$ or $v$ (taking chart transitions into account). This coincides with the classical definition of the 2D motorcycle graph.

For $n=3$ the singularities are curves \cite{liu2018singularity} and the $2$-dimensional isoparametric submanifolds are surfaces $c(s,t)$ on which $\phi(c(s,t)) = (u,v,w)$ is constant in either $u$, $v$, or $w$.
Note that all singular curves are isoparametric curves (Sec.~\ref{sec:seamless}), i.e., they are contained in such isoparametric surfaces, such that the fire starting on different points of a singular curve will spread in common iso-surfaces.
The example in Fig.~\ref{fig:snapshotsSymbolic} illustrates the concept. We discuss the properties of the implied decomposition of $M$ in Sec.~\ref{sec:properties}.

Let us point out an important difference between the 2D and the 3D case: In 2D, the fire front at any time consists of a set of isolated points. Whenever such a point reaches a location already burnt, it dies. This gave rise to the original motorcycle metaphor. In 3D, the fire front is a set of curves (a continuum of points). Such a curve may partially reach burnt terrain, and the remainder proceeds (flowing around the obstacle;  Fig.~\ref{fig:snapshotsSymbolic} center).
The motorcycle metaphor thus, in contrast to the confined brush fire, applies only loosely to the \emph{process} in 3D, but we adopt the name due to the very close analogy in terms of its \emph{results}, the partitions and their properties.
Note that considering the fire front curves as atomic entities instead, that completely stop when any part reaches burnt terrain, would not yield the desired partition properties discussed in the following.

\vspace{-0.1cm}
\subsection{Properties}
\label{sec:properties}

\begin{figure*}
\vspace{0.4cm}
\begin{overpic}[width=0.99\textwidth]{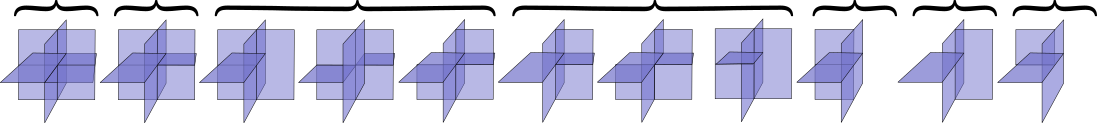}
\put(3,12){(12\,|\,8)}
\put(12,12){(11\,|\,6)}
\put(30.5,12){(10\,|\,4)}
\put(57.8,12){(9\,|\,2)}
\put(75.8,12){(8\,|\,4)}
\put(85,12){(8\,|\,0)}
\put(94,12){(7\,|\,2)}
\end{overpic}
\vspace{-0.2cm}
\caption{All wall configurations that may occur around a single node in a regular region (up to symmetry). The values ($w$\,|\,$c$) specify the numbers of walls $w$ and the numbers of corners $c$ (all of \degn type) incident at the node.
Any other configuration cannot occur in the motorcycle complex because it contains open edges or \degt edges (see Sec.~\ref{sec:blocktypes}).
\label{fig:corners}}
\vspace{-0.25cm}
\end{figure*}

We define the following terminology:
\begin{itemize}
    \item \emph{node}: \tabto{1cm} intersection point of multiple arcs.
    \item \emph{arc}: \tabto{1cm} intersection curve of multiple walls.
    \item \emph{wall}: \tabto{1cm} part of the burnt space bounded by arcs.
    \item \emph{block}: \tabto{1cm} part of $M$ bounded by walls.
\end{itemize}
These entities form the 0-, 1-, 2-, and 3-dimensional cells of a non-conforming generalized cell complex; in contrast to the usual definition of a cell complex, the $k$-cells implied by the brush fire construction are not necessarily homeomorphic to a $k$-ball. We discuss (and resolve) this in the following.

\subsubsection{Block Regularity}

By construction, all singular points of $\phi$ are contained in arcs or nodes. The interior of each block (as well as the interior of each wall) therefore contains only regular points; when restricted to a single block $b$, $\phi\big|_b$~is regular.

\subsubsection{Element Types}
\label{sec:blocktypes}

First, we can observe that blocks have a boundary that is piecewise planar with respect to $\phi$. This is because a block is bounded by walls, and walls are isoplanes of $\phi$. We can establish that these planar pieces (called block facets) meet only in \degn edges (at arcs) and in pairwise \degn corners (i.e., corners where three \degn edges meet). These angles are  to be understood w.r.t. $\phi$.

\paragraph*{Edges} Around singular curves, clearly, isoparametric surfaces emanate in \degn intervals. Therefore, at singularities blocks have \degn edges.
Away from singularities, walls end only where they hit another wall or the model's boundary. In both cases, because both walls and the boundary are (piecewise) iso-parametric (and different iso-planes are orthogonal in $\phi$), \degn edges are formed (cf.~Fig.~\ref{fig:Ts}).

\paragraph*{Regular Corners} 
Corners are formed wherever more than two walls meet in one point. This can be the case at singularities (where walls emanate in the brush fire process) and at points away from singularities where multiple planes meet in the course of the brush fire process.
Fig.~\ref{fig:corners} lists all the possible wall configurations that may occur at such a regular point. Obviously, they are all subsets of the \emph{complete} configuration labeled (12\,|\,8), with the maximum of 12 walls meeting in one point. All other subsets (those not depicted)
\begin{wrapfigure}[5]{r}{0.14\linewidth}
\vspace{-3.0mm}
\hspace{-0.7cm}
\begin{overpic}[width=1.5\linewidth]{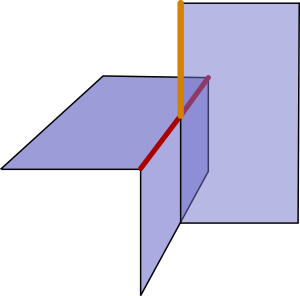}
\end{overpic}
\end{wrapfigure}
contain
an open edge or a \degt edge, as illustrated in the inset in
orange and red. Open edges cannot occur because the brush fire would not have stopped there; \degt edges cannot occur because at least one of the two incident walls would have continued.
 
 \paragraph*{Singular Corners} 
 At a singular point the configuration looks different, and depends on the singularity type (which there are infinitely many of \cite{liu2018singularity}). In any case, however, if a corner would
\begin{wrapfigure}[6]{r}{0.2\linewidth}
\vspace{-3.7mm}
\hspace{-0.7cm}
\begin{overpic}[width=1.35\linewidth]{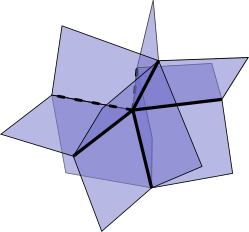}
\end{overpic}
\end{wrapfigure} 
be formed that is not a \degn corner, this would imply there is
an incident block edge that is not a \degn edge. At singular curves (bold black in the inset), however, only \degn edges are formed, and around potential additional regular isolines incident to singular points (dotted) the situation is analogous to the above regular case: open edges and \degt cannot occur, so only \degn edges are possible.

\paragraph*{Facet Types}
Given this restriction to \degn corners, according to the Gauss-Bonnet theorem a block's facet (consisting of one or more walls) must be a disk with 4 corners (a rectangle), or an annulus with no corners. Closed (spherical or toroidal) facets do not occur; they would lack a fire source. A rectangle facet cannot be adjacent to an annulus facet of the same block: at the corners of a rectangle, glued to an annulus at a \degn edge, either a corner in the annulus, or an adjacent \dego edge would be implied, both of which we ruled out. A block thus has either rectangular or annulus facets, not both.

\paragraph*{Block Types}
For a block the Gauss-Bonnet theorem, together with the restriction to \degn corners, implies that it must be either of genus 0 with 8 such corners, or of genus 1 with no corners. 

A block of genus 0 cannot have annulus facets, as they would not form any corners. According to Euler's formula, it must therefore have 6 {(rectangular)} facets. There are only two (structurally distinct) polyhedra with 8 vertices and 6 faces: the cube and the tetragonal antiwedge. Of these only the cube has 
{four-sided facets}. We conclude that if a block is simply-connected, it must be a cuboid (in particular a rectangular cuboid with respect to the $\phi$-metric).

\begin{figure}[tb]
\vspace{-0.3cm}
\includegraphics[width=.99\columnwidth]{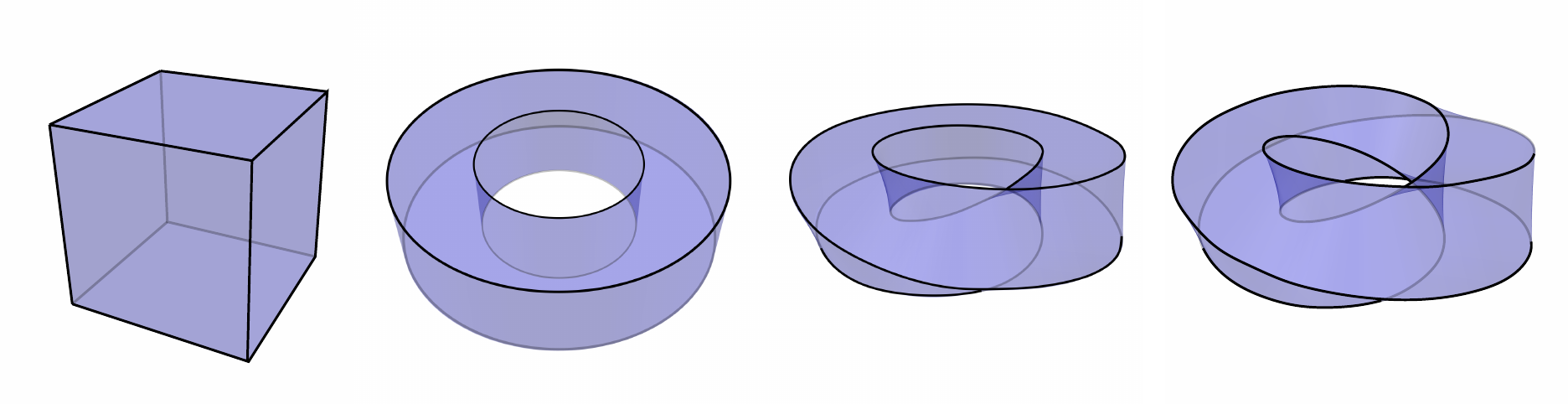}
\vspace{-0.3cm}
\caption{Blocks of the raw motorcycle complex can only be cuboidal or toroidal. Toroidal blocks may have 4, 1, or 2 annulus facets depending on their \emph{twist} (here 0, $\frac{1}{4}$, and $\frac{1}{2}$, respectively).\label{fig:blocktype}}
\vspace{-0.15cm}
\end{figure}

For a block of genus 1, facets must be annuli, so as to not form any corner. Blocks then must be tori, with rectangular cross section. There is an infinite number of structurally different such tori: the \emph{twist} of the torus can be an arbitrary number of quarter turns. In case of twist $k\frac{1}{4}$, the block has 4 facets if $k\!\!\mod 4 = 0$, 2 facets if $k\!\!\mod 4 = 2$, and only 1 facet if $k$ is odd.
Fig.~\ref{fig:blocktype} illustrates these types of blocks that initially can occur in the motorcycle complex.

\begin{wrapfigure}[5]{r}{0.17\linewidth}
\vspace{-3.2mm}
\hspace{-0.7cm}
\begin{overpic}[width=1.395\linewidth]{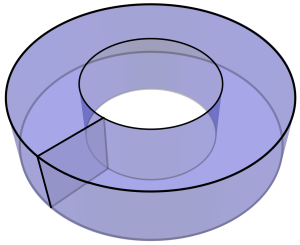}
\end{overpic}
\end{wrapfigure}
Any genus 1 block, regardless of its twist, can be turned into a (self-adjacent) genus~0 block by introducing one additional wall that cuts it. Using this modification (Sec.~\ref{sec:impl:split}), a pure cuboid block complex is obtained in any case, avoiding the need for further special case handling in subsequent operations.

\subsection{Reducibility}
\label{sec:reducible}

Just as in the 2D case (Sec.~\ref{sec:mc2d}), we cannot expect the resulting motorcycle complex to describe a globally minimal (i.e., smallest) partition with the desired properties (cuboidal, regular, aligned). It is worthwhile considering the aspect of local minimality, though. To this end we define the following (for the 3D and 2D cases):

\begin{definition}[Reduction]
The operation of merging two adjacent $n$-cells of a cuboid (quadrilateral) cell complex into one $n$-cell that is cuboid (quadrilateral) we call a \emph{reduction}. A complex that allows for no such reduction of any two $n$-cells we call \emph{irreducible}. \label{def:reduce}
\end{definition}
Two $n$-cells are adjacent if they share an $(n\!-\!1)$-cell (an arc in the 2D case, a wall in the 3D case). 
If they can be merged in a reduction, we say the shared cell is \emph{removable}.

A restricted notion is that of \emph{regular-irreducibility}, defined via \emph{regular-reductions} that merge cells only across \emph{regular-removable} arcs or walls not incident to any singular node or arc. This notion is relevant for use cases (such as in Sec.~\ref{sec:spline}) that require singularities lie only at block edges, not in the interior of block facets.

\begin{figure}[tb]
\begin{overpic}[width=.99\linewidth]{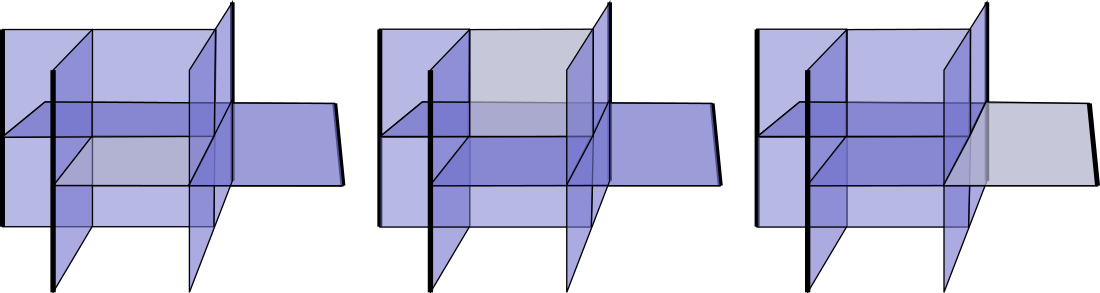}
\end{overpic}
\vspace{-0.1cm}
\caption{Illustration of \emph{reducibility}, based on the example from Fig.~\ref{fig:snapshotsSymbolic}. Left: the grey wall can be removed, merging two cuboid cells into one that is still cuboid. Center: the grey wall cannot be removed as it would yield a non-cuboid block. Right: the grey wall is not regular-removable as it is incident to a singular arc (that would end up lying inside the merged block's facet).\label{fig:reducible}}
\vspace{-0.1cm}
\end{figure}

Under a general position assumption on the location of singularities, the standard 2D motorcycle graph (while reducible) is regular-irreducible; only when motorcycles meet in a frontal manner there may be options for regular-reduction. 
As demonstrated in Fig.~\ref{fig:reducible}, the situation is different in the 3D case: even regular-irreducibility is not a given, the brush fire result commonly contains regular-removable walls.
The underlying reason is related to the discussion at the beginning of Sec.~\ref{sec:themc}: while motorcycles in 2D are points, and collisions with traces are isolated instantaneous events, in 3D the more complex brush fire that forms a wall may stop in one place while proceeding in another.

\paragraph*{Wall Retraction} We therefore propose to subsequently reduce the result of the brush fire process to a locally minimal, i.e., either regular-irreducible or irreducible state, as desired. To this end, we perform \emph{wall retraction}: (regular-)removable walls are greedily removed, ordered by their parametric distance from their origin. Intuitively, this can be interpreted as retracting fire walls in places where they have spread unnecessarily far in the brush fire expansion.
hile it would be conceptually attractive to determine and avoid this redundancy already during the expansion process, we are not aware of a possible strategy to steer the process accordingly. Practically, however, the overhead due to the reduction happening after the fact is benign (see experiments in Sec.~\ref{sec:results}).

Where distinction is necessary, we refer to the non-reduced brush fire result as \emph{raw motorcycle complex}, while \emph{motorcycle complex} is generally meant to refer to the reduced version (with additional walls inserted in rare toroidal cells, Sec.~\ref{sec:blocktypes}). In Sec.~\ref{sec:impl} we describe the construction as well as the reduction process in detail.

\paragraph*{Remark (Base Complex Reduction)} One could start from the base complex and apply reductions until an irreducible minimum is achieved. However, the base complex
can be very large, hampering practical construction, and, according to our experiments reported in Sec.~\ref{sec:results}, retraction starting from the base complex commonly ends up in worse local minima, i.e., complexes of larger size.

\paragraph*{Remark (Sparse Serial Construction)}
In the 2D case, a not only regular-irreducible but fully irreducible motorcycle graph can be obtained right away by not tracing motorcycles simultaneously but serially, and omitting motorcycles whose neighbors around a singularity have already been traced \cite[\S7]{eppstein2008motorcycle}. This strategy can be applied in the 3D setting as well, as we detail in the supplementary material (part A). However, the reported experiments show that this serial strategy commonly leads to more complex results than wall-retraction applied to the standard simultaneous strategy; we therefore focus on the latter in the following.

%% file: text/05implementation.tex
\section{Implementation}
\label{sec:impl}

We describe two implementations, one to compute a motorcycle complex of a hexahedral mesh (primarily as an intuitive entry) and one to compute a motorcycle complex of a seamless parametrization on a tetrahedral mesh. In the former case we can exploit that all relevant isosurfaces are available explicitly as facets of hexahedral elements, resulting in a {discrete (combinatorial rather than geometric)} algorithm; in the latter case isosurfaces arbitrarily cross mesh elements in a continuous manner, requiring additional efforts.

\subsection{Mesh-based}
\label{sec:alg:mesh}

In this case the input is a hexahedral mesh, consisting of vertices, edges, facets, and hexes. {Implicitly, it has a natural seamless para-metrization, mapping hexes to unit cubes.} Interior edges are \emph{singular} if their number of incident hexes is different from 4; boundary edges if it is different from 2.
For a regular edge $e$ and an incident facet $f$ let $\opp(f,e)$ denote the facet incident to $e$ not incident to a common hex with $f$; for a boundary edge it may not exist.
For an edge $e$, $F_e$ denotes the set of incident interior facets.

The algorithm makes use of a priority queue $Q$ of $(e,f,d)$ tuples, each with an edge $e$, a facet $f$, and a distance $d\in \mathbb{N}$. Queue elements are ordered by $d$, smallest first.

The condition $\alive(e)$ (line \textbf{1}) is true iff at most two facets incident to $e$ are tagged or $e$ is singular.
This means that an edge that has already been crossed will not be crossed again (in orthogonal direction).
This implies that ties (two fire walls reaching an edge orthogonally with the same distance $d$) are broken arbitrarily. 
Note that even if the tie is broken differently on neighboring edges, the resulting partition will be structurally valid (as if both fire fronts had continued), i.e., there is no need for global coordination.

\begin{algorithm}
\SetAlgoLined
\DontPrintSemicolon
\lForEach(\tcp*[f]{ignite}){singular $e$}{$Q$.push$\{(e,f,0) \mid f\!\in\! F_e\}$}
\While{$Q$ non-empty}{
$(e,f,d) \gets Q$.pop()\;
\nl  \If(\tcp*[f]{not crossing burnt terrain}){$\alive(e)$}
{tag $f$\tcp*{mark facet as burnt}
\ForEach{regular interior edge $e' \neq e$ incident to $f$}
{\lIf(\tcp*[f]{spread}){$\opp(e',f)$ is not tagged}{$Q$.push$(e', \opp(e',f), d+1)$}}
}}
\lForEach{boundary facet $f$}{tag $f$} 
\caption{Motorcycle Complex of Hexahedral Mesh\label{alg:mesh}}
\end{algorithm}
\vspace{-0.1cm}
Once the algorithm terminates, the union of all tagged facets form the walls of the raw motorcycle complex, partitioning the hexahedral mesh into blocks $B_i$, each consisting of $m_i\times n_i\times o_i$ hexes for some $m_i,n_i,o_i\in\mathbb{N}$. The explicit structure and connectivity of the motorcycle complex is then easily discovered by exploiting the connectivity of the underlying hexahedral mesh.

\vspace{-0.1cm}
\subsection{Parametrization-based}
\label{sec:impl:param}

Here the input is a tetrahedral mesh, consisting of vertices, edges, facets, and tets, equipped with a seamless parametrization. 
In contrast to the algorithm in Sec.~\ref{sec:alg:mesh} here we cannot simply walk along the faces of the mesh: the isosurfaces relevant for the motorcycle complex do not coincide with the tetrahedral mesh's facets, but cross its tets arbitrarily. We thus need to perform the brush fire expansion through the interior of tets. Inside each tet the situation can furthermore be highly complex, with multiple fire walls meeting in arbitrary configurations; essentially, within each tet a separate Euclidean 3D motorcycle complex problem is to be dealt with.

We can simplify implementation significantly by refining the mesh on the fly while spreading the fire, so as to have it coincide with facets of the mesh. This simplifies not only the propagation process, but also the representation of the motorcycle complex and the final discovery of its  structure and connectivity.
The following algorithm spells out this process. Notice the close analogy to Alg.~\ref{alg:mesh}, extended to perform and deal with the refinement of the mesh.

\begin{algorithm}
\SetAlgoLined
\DontPrintSemicolon
\ForEach(\tcp*[f]{ignite}){singular $e$}{
\ForEach{tet $t$ incident on $e$}{
\nl    \lIf{$f \gets \spli(e,t)$}{$Q$.push$(e,f,0,n)$}}}
\While{$Q$ non-empty}{
$(e,f,d,n) \gets Q$.pop()\;
\If(\tcp*[f]{not crossing burnt terrain}){$\alive(e)$}
{tag $f$\tcp*{mark facet as burnt}
\ForEach{regular interior edge $e' \neq e$ incident to $f$}
{\ForEach{tet $t$ incident on $e'$}{
\nl  \If{$f' \!\!\!\gets\! \spliopp(e'\!,t,f)$ $\!\wedge\!$ $f'\!$ not tagged}{
\nl  $Q$.push$(e', f', d+\exte(e,e',n), \tau n)$}}
}}} 
\lForEach{boundary facet $f$}{tag $f$}
\parbox{\linewidth}{\caption{Motorcycle Complex of Seamless Parametrization\label{alg:param}}}
\end{algorithm}

\vspace{-0.2cm}
\subsubsection{Mesh Refinement} 
\begin{figure}[b]
\vspace{-0.1cm}
    \newcommand*{\xMin}{-2}%
    \newcommand*{\xMax}{2}%
    \newcommand*{\yMin}{-1}%
    \newcommand*{\yMax}{1}%
    \begin{tikzpicture}
        \node[inner sep=0pt] at (-3,0)
        {\begin{overpic}[width=.45\linewidth]{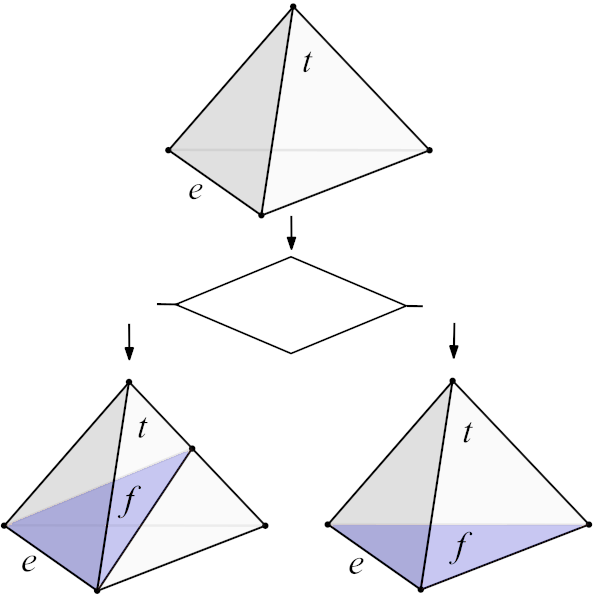}
        \put(17.3,48){\footnotesize no}
        \put(71.8,47.9){\footnotesize yes}
        \put(37.8,47.7){\tiny iso-facet?}
        \end{overpic}};
        \node[inner sep=0pt] at (1.5,-0.68)
        {\includegraphics[width=.45\linewidth]{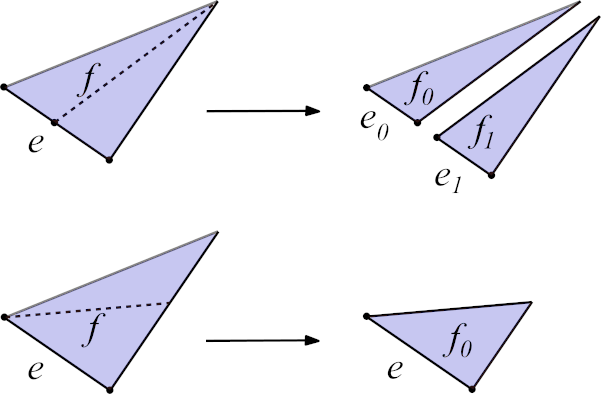}};

\node[] at (-4.6,1.55) {(a)};
\node[] at (-0.25,1.55) {(b)};

    \end{tikzpicture}
    \vspace{-0.0cm}
    \caption{a) Splitting a tetrahedron along an isoplane using $\spli(e,t)$. The returned new facet $f$ is marked blue; the special case of an existing iso-facet being returned is also shown. b)~Updating a queue entry $(e,f,\cdot,\cdot)$ when $e$ gets split (top), and when $f$ but not $e$ gets split (bottom).}
    \label{fig:split}
    \vspace{-0.1cm}
\end{figure}

The method $\spli(e,t)$ (line \textbf{1}) performs the following: if there is a parametric iso-plane that contains~$e$ and intersects the opposite edge $e'$ of $t$ at a point $p$, the edge $e'$ is split at $p$, introducing a new vertex $v$ and splitting all incident tets, and the new iso-facet formed by $e$ and $v$ is returned. Otherwise, if any of the two facets of $t$ incident on $e$ is an iso-facet, it is returned.
An iso-facet is a facet constant in one of the $\phi$-parameter values ($u$, $v$, or~$w$).
These cases are illustrated in Fig.~\ref{fig:split}a.

The method $\spliopp(e,t,f)$ (line \textbf{2}) behaves as $\spli(e,t)$, but considers only iso-facets aligned with $f$ (same constant parameter, taking transitions into account) except $f$ itself.

Additionally, whenever such a split is performed, affected edges and facets in the queue need to be updated. When an edge $e$ is split, each queue entry $(e,f)$ needs to be replaced by two entries $(e_0,f_0)$ and $(e_1,f_0)$, with sub-edges $e_0$, $e_1$ and sub-facets $f_0, f_1$ (see Fig.~\ref{fig:split}b top). When a facet $f$ is split, but not the edge $e$ of an entry $(e,f)$, it is replaced by $(e,f_0)$, where $f_0$ is the sub-facet incident on $e$ (see Fig.~\ref{fig:split}b bottom).
A more efficient (slightly more involved) implementation alternative is to postpone these updates (avoiding the queue update). We keep a binary forest that records the facet split hierarchy: for each facet that gets split, a record of the two resulting sub-facets it kept. When an entry with facet $f$ and edge $e$ is popped from the queue but $f$ does not exist in the mesh anymore (because it was split), we look up its two children in the hierarchy. Either one or two of these has an edge that is a sub-edge of $e$ (the two cases in Fig.~\ref{fig:split}b). We push the children \emph{with} an $e$-sub-edge into the queue and continue. This may proceed recursively, until the sub-elements currently present in the mesh are reached.

{A further modification over Alg.~\ref{alg:mesh} is necessary for Alg.~\ref{alg:param}: The queue is ordered by $d$ only secondarily; primarily, queue entries with $e$ not lying in an original mesh facet are given priority. This ensures that once the brush fire front has entered the space of an original tet, it (atomically) proceeds through this space entirely (i.e. through all refinement-induced sub-tets). This prevents potentially infinite alternating split sequences that could occur when multiple fire walls were spreading inside the same original tet.}

\begin{figure*}[t]
  \includegraphics[width=0.99\linewidth]{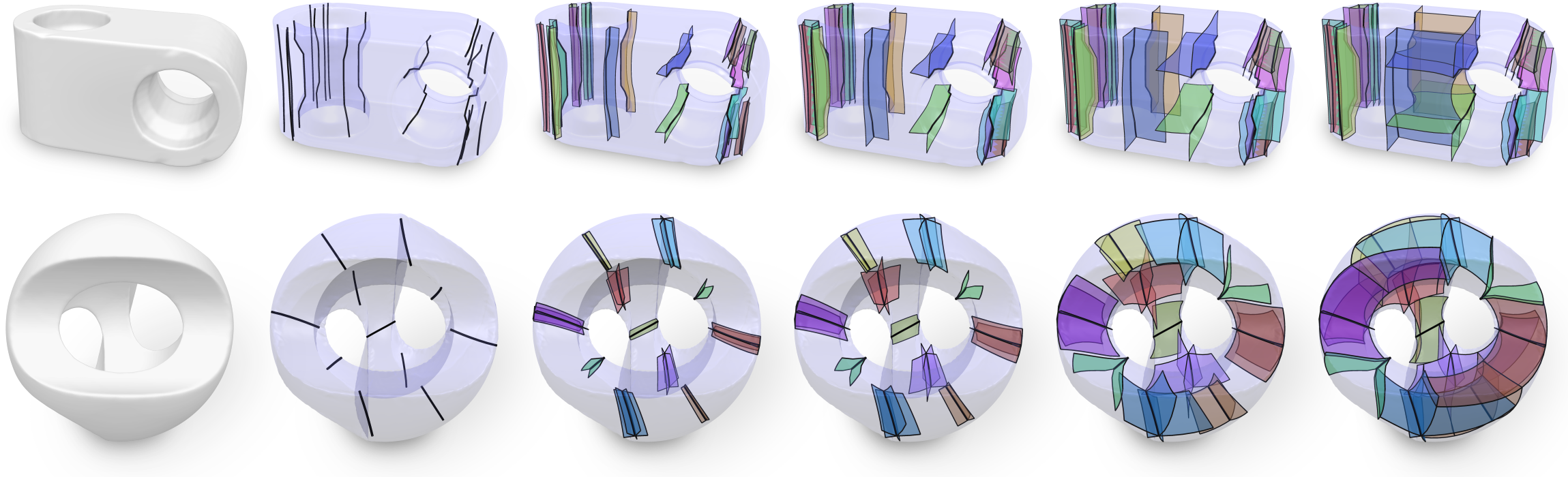}
  \vspace{-0.35cm}
  \caption{Snapshots of the motorcycle complex construction algorithm (left to right) in seamless parametrizations of two example objects. The black curves in the second column are the parametrization's singularities, which spawn the fire walls that partition the object's interior. {For visual clarity, fire fronts are shown as smooth curves here; the supplementary video shows the raw propagation process in more detail.}\label{fig:snapshots}}
  \vspace{-0.15cm}
\end{figure*}

\vspace{-0.1cm}
\subsubsection{Distance Tracking} Compared to the 
algorithm in Sec.~\ref{sec:alg:mesh}, in which propagation distances can be increased discretely in unit steps, additional attention is required here.
The function $\exte(e,e',n)$ (line \textbf{3}) is defined as follows:
Let $p_e$ be the end point of $e$ for which $n^T\phi(p_e)$ is minimal; then we define $\exte(e,e',n) = n^T(\phi(p_{e'})-\phi(p_e))$ (w.r.t.~the coordinate chart of facet $f$).
Here $n$ is a unit axis-aligned vector (conceptually the parametric direction of propagation). In the initialization (line \textbf{1}) it is orthogonal to the singular edge $e$ and contained in $f$. During propagation it is transformed using $\tau$ (line \textbf{3}), the chart transition between $f$ and $f'$ around $e'$; in this we assume each facet is arbitrarily associated with one of its two adjacent tets, adopting its chart coordinate system.

Note that this implementation performs propagation in a discrete facet-by-facet manner, i.e., wall collisions are not handled in a continuous manner. In particular, this can lead to non-minimal results. But as non-minimality is an inherent property either way (Sec.~\ref{sec:reducible}), and as we are therefore going to reduce the resulting complex anyway, the significant added complexity of a continuous collision resolution would be unlikely to be justified in practice.

Fig.~\ref{fig:snapshots} illustrates the algorithm on examples.
Source code for this propagation process is available in the supplementary material.

\subsection{Torus Splitting}
\label{sec:impl:split}

In order to turn occasional genus 1 blocks (cf.~Fig.~\ref{fig:blocktype}) into cuboid blocks, one simply detects these (by counting corners) and starts a new brush fire inside the block from an arbitrary point on one of its arcs, confined to the iso-plane that is orthogonal to the two walls incident at that point. This yields an additional wall, cutting the toroidal block to a (self-adjacent) cuboid block.

\subsection{Reduction}

Following Def.~\ref{def:reduce}, a wall can be removed from the cell complex if the union of its two adjacent blocks is again cuboid and, optionally, it is not incident to a singular arc. This is easily determined: for each of the four arcs surrounding a wall, check that
\begin{itemize}
    \item the two wall-adjacent blocks form a \degn edge (rather than a \dego edge) each at that arc,
    \item the two wall-adjacent blocks are actually distinct,
    \item and optionally: the arc is regular.
\end{itemize}
Removable walls are queued up, sorted by parametric distance to their origin. This distance is available as value $d$ per facet~$f$ during Algorithm~\ref{alg:param} and stored accordingly. We then greedily remove removable walls, starting with the farthest. Whenever a wall is removed, (some of) its arcs may become trivial in the sense that only two incident walls are left; these arcs vanish and the two incident walls are merged. The removability status of all adjacent walls is then retested and the queue updated accordingly.

%% file: text/06sanitization.tex
\section{Parametrization Sanitization}
\label{sec:sanity}

Seamless parametrizations are commonly obtained through numerical optimization routines \cite{Nieser:2011}.
This involves inaccuracies due to limited precision. The resulting parametrizations therefore commonly are not exactly seamless on a numerical level. This bears some potential of leading to inconsistencies in the construction of the motorcycle complex. For the 2D case, this issue was discussed in detail in previous work \cite{ebke2013qex,mandad2019exact}. The latter article proposes a method that transforms a nearly seamless parametrization of a triangle mesh into one that is truly seamless---while preserving its singularities and boundary alignment. In this section we describe a generalization to the volumetric case on tetrahedral meshes. This enables the safe application of the motorcycle complex algorithm on the resulting truly seamless volumetric parametrization.

\subsection{Background: 2D Case}

We briefly recapitulate the 2D case, focusing on the differences and referring to the original paper for a complete overview.

The overall constraint system for seamless parametrization consists of chart transition constraints across all non-boundary edges, alignment constraints along the boundary and feature edges, and possibly further constraints like cycle or connection constraints.

\paragraph*{Exact Constraint Satisfaction.} To obtain an exact solution to the constraint system, close to the given almost-seamless parametrization, \cite{mandad2019exact} propose to first separate the variables into two sets---\emph{implied} and \emph{free}---by converting it into integer reduced row echelon form. This can be done without numerical error, confined to the integer domain. This turns the system into upper triangular form, such that all the {implied} variables are expressed as linear combinations of {free} variables with solely rational coefficients.
The {free} variables can then be chosen in such a way that the {implied} variables can be computed and represented without error using standard floating point arithmetic.
To this end, the {free} variables are initialized according to the values in the given almost-seamless parametrization, but then slightly altered and quantized such that they are divisible by everything they will be divided by in the linear combinations. In this way, the method ensures that ultimately all variables are standard floating point numbers while exactly satisfying all constraints.

This approach is generic and in principle applicable to any homogeneous constraint system (including 3D seamlessness constraints). 
However, the above constraints form a large system with a size of the same order as the mesh; variables are the $(u,v)$-parameters of the mesh's vertices. In such cases the approach is impractical. \cite{mandad2019exact} showed how a simplified core system (over only certain \emph{sector} variables of particular \emph{node} vertices) can instead be considered, drastically reducing the effective system size.
Via generalization, we follow an analogous path for the 3D case.

\subsection{3D Case}

The seamless parametrization $\phi$ consists of linear maps $\phi^t: t\rightarrow \mathbb{R}^3$ per tetrahedron $t$, related across the tetrahedras' facets via transition functions (cf.~Sec.~\ref{sec:seamless}). The transition function across a facet in one direction is the inverse of that across it in the opposite direction. 

For our purpose, we are interested in constraints for seamless transitions and boundary alignment being satisfied exactly.

\paragraph*{Transition Constraints.} Seamlessness can be imposed by requiring for each edge $ab$ of a facet (with \emph{intended} transition function $\pi_{st}$) between two tetrahedra $s$ and $t$:
\begin{equation}
\label{eq:transition}
\phi^t(b) - \phi^t(a) = \pi_{st} (\phi^s(b) - \phi^s(a)).
\end{equation}
Note that only the rotational (not the translational) part of the rigid transformation $\pi_{st}$ matters in this formulation, as it is applied to vectors rather than points.

\paragraph*{Alignment Constraints.} For boundary alignment of a facet $f$ of a tetrahedron $t$, one requires that one particular of the parametrization's three components $(u,v,w)$ is constant along each edge $ab$ of~$f$:
\vspace{-0.3cm}
\begin{equation}
\label{eq:alignment_constraint}
\phi^t(b)|_k = \phi^t(a)|_k,
\end{equation}
where $k$ is 0, 1, or 2, depending on the respective component. To ensure boundary alignment, such a constraint is in effect for all boundary edges.

\subsubsection{Terminology}

A facet for which the intended transition function is not identity we call a \emph{cut facet}. The union of all cut facets forms the \emph{cut set}.
We call an edge a \emph{cut edge} if one of the following holds:
\begin{itemize}
    \item it is a singularity edge,
    \item it is incident to one or to more than two cut facets,
    \item it is a boundary edge and incident to at least one cut facet, or, its incident boundary facets have different alignment.
\end{itemize}

For the matter of this section (for consistency with previous work) we will use the term \emph{node} with a different meaning than in the context of the motorcycle complex in Sec.~\ref{sec:properties}. As this section deals with an orthogonal matter and is not concerned with the motorcycle complex, no ambiguities are caused.

We refer to a vertex as \emph{node} if (i) it is incident to one or to more than two cut edges, or (ii) it is a boundary vertex incident to a non-boundary cut edge. A connected set of cut edges bounded by \emph{nodes} form a \emph{branch}.
All these branches together partition the cut set and the mesh boundary into pieces we call \emph{sheets}. 
Each sheet is an orientable 2-manifold surface (with one or multiple boundary loops) and is either a \emph{cut} sheet or an \emph{align} (i.e.~boundary) sheet. 
Note that within each cut sheet, the transition function is constant, and within each boundary sheet, the aligned component ($k$ in Eq.~\eqref{eq:alignment_constraint}) is constant.
Around each vertex, the cut facets partition the tetrahedral mesh into \emph{sectors}, such that all incident tetrahedra within a sector share parametrization values at the vertex.

\subsubsection{Simplified Constraint System}
\label{sec:simplesystem}

The overall constraint system to be satisfied consists of transition constraints across all mesh facets and alignment constraints over boundary facets.
In the supplementary material (part B) we show that the sub-system concerned with the non-node vertices can be converted into triangular form, similar to the 2D case \cite{mandad2019exact}. It is therefore sufficient to deal with a small core system involving only the variables associated with nodes---a number proportional to the complexity of the singularity structure of $\phi$ (assuming a sensible cut choice), rather than to the size of the mesh.
The proper parametrization values for non-node vertices can easily be computed from the result, in a back-substitution-like manner, afterwards.

More precisely, for this simplified system, we need to consider one $\bm{u}$-variable per node sector. Note that $\bm{u}=(u,v,w)$ has three components. 
For each sheet, we mark one of its nodes as \emph{base} node.
In rather rare cases there may be branches which are circular; on these we consider two arbitrary vertices as additional nodes. And there may be loop branches, starting and ending at the same node; on these we consider one arbitrary vertex as additional node. In this way each sheet has at least two nodes on it.

\begin{wrapfigure}[7]{r}{0.2\linewidth}
\vspace{-4.0mm}
\hspace{-0.6cm}
\begin{overpic}[width=1.1\linewidth]{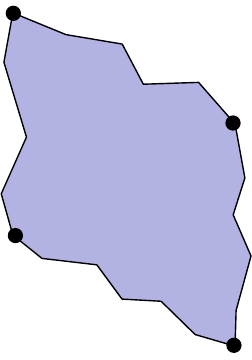}
 \put(5,39){$\bm{u}^+_{0}$}
 \put(5,81){$\bm{u}^+_{3}$}
 \put(13,95){$\bm{u}^-_{3}$}
 \put(53,12){$\bm{u}^+_{1}$}
 \put(69,12){$\bm{u}^-_{1}$}
 \put(52,55){$\bm{u}^+_{2}$}
 \put(70,67){$\bm{u}^-_{2}$}
\end{overpic}
\end{wrapfigure}
For each sheet (with its marked base node) every other node on it will contribute one equation---either transition or alignment---depending on
the type of the sheet.
More specifically, given a cut sheet with transition function $\pi$ and having $n$ nodes with sector variables $\bm{u}^\pm_0, \bm{u}^\pm_1, \bm{u}^\pm_2, \ldots, \bm{u}^\pm_{n-1}$ ($\pm$ denoting sector variables on the two sides, front and back, of the sheet; see inset figure), the equation corresponding to the $i$-th node ($0 < i < n$) will be: 
$
\bm{u}^+_i - \bm{u}^+_0 = \pi (\bm{u}^-_i - \bm{u}^-_0).
$
Similarly, for an align sheet we set up an equation per node: 
$
\bm{u}_i|_k =  \bm{u}_0|_k
$ (no $\pm$-distinction on the boundary).

This simple system is then solved as in \cite[\S5.3.1]{mandad2019exact} ensuring that all the constraints are satisfied while conveniently remaining in the floating point domain.
From the result, the precise transition function (including its translational component) of each cut sheet, and the precise constant parameter value of each align sheet is determined (i.e., can be read from the new values at nodes). The parametrization values of each non-node sheet vertex can then easily be updated to match this: fix the $\bm{u}$-value in one sector (after appropriate precision truncation, and possibly taking the determined alignment value into account), and propagate it to the other sectors of this vertex using the determined transition functions. At non-node vertices on \emph{singular} branches a little extra care is required: the composed transitions around the branch imply fixpoint parameters \cite[§3.2]{ebke2013qex}; these need to be chosen, so as to keep the propagation consistent across all sectors.

Afterwards, the parametrization is exactly seamless and exactly boundary-aligned, such that the motorcycle complex construction algorithm from Sec.~\ref{sec:impl:param} can be applied safely.

\vspace{-0.05cm}

%% file: text/07results.tex
\section{Results}
\label{sec:results}

In the following we evaluate the characteristics of the motorcycle complex, in particular in comparison to the base complex, as well as the proposed algorithms (mesh-based and parametrization-based) for its construction. The supplementary video provides several animated impressions of the process.
Hexahedral mesh visualizations are rendered using \cite{bracci2019hexalab}.

\begin{table}[t]
\rowcolors{1}{}{backgray}
\begin{filecontents*}{tables/StatsMay18hex_shrunk.csv}
Model;BC;BCR;MCraw;MCplus;MCpluspercent;Tpercent;MC;MCpercent
example 3;406136;67828;9087;5780;1.4\%;42\%;2877;0.7\%
example 1;74331;11385;3137;2248;3.0\%;15\%;1123;1.5\%
example 2;3253;678;233;195;6.0\%;14\%;87;2.7\%
dragon-hex;12488;2959;979;724;5.8\%;36\%;357;2.9\%
gargoyle;7563;1967;720;546;7.2\%;38\%;257;3.4\%
anc101 a1;12336;3118;1359;846;6.9\%;45\%;460;3.7\%
fertility-hex;2002;548;221;189;9.4\%;27\%;76;3.8\%
pegasus-hex;9745;2415;1035;729;7.5\%;36\%;374;3.8\%
kiss hex;5019;1194;543;385;7.7\%;39\%;200;4.0\%
anc101;5009;1283;609;347;6.9\%;39\%;207;4.1\%
impeller stresstest;878;176;184;124;14.1\%;24\%;37;4.2\%
armadillo hex-a;5960;1491;680;516;8.7\%;34\%;266;4.5\%
armadillo hex-b;3265;820;396;296;9.1\%;31\%;147;4.5\%
;;;\textcolor{black}{\Large$\vphantom{\int^0}${\bfseries\small$\smash[t]{\vdots}$}\vspace{-0.1em}};;;;;
example 5;1;1;1;1;100\%;0\%;1;100\%
\end{filecontents*}
\csvreader[separator=semicolon, before reading=\begin{adjustbox}{max width=\columnwidth}, after reading=\end{adjustbox},tabular=lrrrrrrrr, head to column names=true,
table head =
\bfseries Model & \bfseries BC &  \textcolor{gray}{BC\textsuperscript{--}} & \textcolor{gray}{raw} & \bfseries \!MC\textsuperscript{+}\!\! & \!\!\bfseries $\frac{\text{MC\textsuperscript{+}}}{\text{BC}}$ & \textcolor{gray}{\;\;\;T\;\;} & \bfseries MC & \!\!\bfseries $\frac{\text{MC}}{\text{BC}}$ \vspace{1mm}\\,
table foot =,
]{tables/StatsMay18hex_shrunk.csv}{}
{{$\vphantom{\int^0}$\textsc{\Model}}\!\!\! & \!\BC & \textcolor{gray}{\BCR} & \textcolor{gray}{\MCraw} & \MCplus & \!\MCpluspercent & \textcolor{gray}{\Tpercent} & \MC & \!\MCpercent}
\caption{Statistics on a dataset of
{hexahedral meshes} (full table in supplementary material). Reported are numbers of blocks in the base complex (BC), reduced base complex (BC\textsuperscript{--}), raw motorcycle complex (raw), reduced motorcycle complex with preserved singularity-adjacent walls (MC\textsuperscript{+}), and fully reduced motorcycle complex (MC) -- ordered by complexity of MC relative to BC. It can be observed that the raw MC is typically larger than the MC\textsuperscript{+} (or MC) by a factor of around 1.4 (or 2.9) only, i.e., construction overhead over a hypothetical direct construction of the final MC is benign. Furthermore, notice that the fully reduced BC\textsuperscript{--} is generally significantly larger 
than the fully reduced MC (see the remark in Sec.~\ref{sec:reducible}). On average one third of the MC's arcs are T-arcs (T).\label{fig:stats}}
\end{table}

\vspace{-0.1cm}
\subsubsection*{Mesh-based Algorithm}

We take a dataset of 261 all-hex meshes, collected by \cite{bracci2019hexalab}, generated by a variety of hexahedral meshing approaches, e.g. \cite{AMUCIP,PHOVER,AMuSF,AMGvVPD}, 
and apply our mesh-based algorithm (Sec.~\ref{sec:alg:mesh}).
Table~\ref{fig:stats} provides statistics on the results, including the motorcycle partitions' size before and after reduction (Sec.~\ref{sec:reducible}). The full table is available in the supplementary material.

An interesting comparison is with respect to the standard base complex. We include the corresponding statistics in Table~\ref{fig:stats}. As can be observed, the motorcycle complex is often simpler by a large factor (here up to $140\times$). 
Besides having an obvious positive effect on construction cost,
the motorcycle complex offers benefits on the application side, as demonstrated in Sec.~\ref{sec:example}.
Splitting of toroidal blocks (Sec.~\ref{sec:impl:split}) occurred in 13 of the models, a total of 48 times.

\begin{table}[t]
\rowcolors{1}{}{backgray}
\begin{filecontents*}{tables/StatsMay18algohex.csv}
Model;NumTetsK;BCblocks;BCwallsK;BCtetsK;BCtracingTime;BCdiscoveryTime;BCtime;MCblocks;MCrblocks;MCwallsK;MCtetsK;MCtracingTime;MCdiscoveryTime;MCreductionTime;MCtime;blocksPercent;blocksFactor;wallsFactor;tetsFactor;timeFactor;tracingFactor;discoveryFactor
rockerarm;97;2446;1680;4743;434.1;115.5;549.6;217;78;154;521;27.9;11.1;1.9;40.8;3;31.4;10.9;9.1;13.5;15.6;10.4
armadillo;163;3110;901;2623;214.3;60.2;274.5;392;132;260;867;46.4;16.8;1.6;64.8;4;23.6;3.5;3.0;4.2;4.6;3.6
joint;42;205;149;454;27.6;9.7;37.3;54;15;56;199;9.5;3.6;0.6;13.7;7;13.7;2.7;2.3;2.7;2.9;2.7
kitten;30;208;81;246;15.2;5.6;20.7;77;19;49;162;8.3;3.4;0.7;12.4;9;10.9;1.6;1.5;1.7;1.8;1.6
broken bullet;20;44;20;74;3.5;1.3;4.8;25;5;13;54;2.1;0.9;0.1;3.1;11;8.8;1.6;1.4;1.6;1.7;1.5
sculpture;20;108;59;184;10.1;3.6;13.7;27;13;17;68;2.8;1.2;0.1;4.0;12;8.3;3.5;2.7;3.4;3.6;3.0
fandisk;46;128;87;284;15.4;5.2;20.6;43;19;38;153;6.6;2.5;0.3;9.4;15;6.7;2.3;1.9;2.2;2.3;2.1
bone;54;87;53;193;9.1;3.5;12.6;57;15;45;171;7.4;3.0;0.8;11.2;17;5.8;1.2;1.1;1.1;1.2;1.2
camille hand;103;142;103;377;18.6;7.5;26.1;75;26;74;299;12.5;5.6;1.2;19.3;18;5.5;1.4;1.3;1.3;1.5;1.3
cylinder;12;26;44;133;7.6;2.7;10.4;11;5;14;52;2.4;0.9;0.1;3.4;19;5.2;3.1;2.5;3.0;3.2;2.9
sphere;19;7;6;34;0.9;0.4;1.4;7;2;6;34;0.9;0.4;0.1;1.4;29;3.5;1.0;1.0;1.0;1.0;1.0
cube sphere;11;10;4;22;0.6;0.3;0.9;10;4;4;22;0.6;0.3;0.0;0.9;40;2.5;1.0;1.0;1.0;1.0;1.0
tetrahedron;14;4;2;18;0.3;0.2;0.5;4;2;2;18;0.3;0.2;0.0;0.6;50;2.0;1.0;1.0;1.0;1.0;1.0
fanpart;5;5;2;11;0.3;0.1;0.5;5;3;2;11;0.3;0.1;0.0;0.5;60;1.7;1.0;1.0;1.0;1.0;1.0
prisma;21;3;3;30;0.6;0.4;0.9;3;2;3;30;0.5;0.4;0.0;0.9;67;1.5;1.0;1.0;1.0;1.0;1.0
\end{filecontents*}
\csvreader[separator=semicolon, before reading=\begin{adjustbox}{max width=\columnwidth}, after reading=\end{adjustbox},tabular=lrrrrrrrr, head to column names=true,
table head =
    \bfseries Model & 
    \;\;\;\;\;\;\textcolor{gray}{tets} &
    \bfseries \;\;\;\;BC & \bfseries MC & 
    \!\!\bfseries $\frac{\text{MC}}{\text{BC}}$ & 
    {\;\textcolor{gray}{facets}}\; & 
   \;\;\;\,trace & build & \!reduce\!
    \vspace{1mm}\\,
table foot =,
]{tables/StatsMay18algohex.csv}{}
    {$\vphantom{\int^0}$\textsc{\Model}\!\! & 
    \textcolor{gray}{\NumTetsK\,K} & 
    \BCblocks & \;\MCrblocks & 
    \,\blocksPercent\% & 
    \textcolor{gray}{\MCwallsK\,K} &
    \MCtracingTime\,s &
    \MCdiscoveryTime\,s &
    \MCreductionTime\,s}
\caption{Statistics on a dataset of seamless parametrizations of tetrahedral meshes. Columns show the number of tetrahedra in the input meshes (\emph{tets}), the number of triangular mesh facets tagged by Alg.~\ref{alg:param} (\emph{facets}), and the time spent in the three algorithmic steps (tracing the fire walls, building a graph representation of the complex (including torus splitting), and wall retraction for reduction). Notice that, similar to Table~\ref{fig:stats}, the BC again has up to $30\times$ as many blocks as the MC of the same model.
\label{fig:statparam}}
\end{table}

\subsubsection*{Parametrization-based Algorithm}

We apply the parametrization-based algorithm (Sec.~\ref{sec:impl:param}) to seamless parametrizations on tetrahedral meshes, generated using frame field guided parametrization, i.e., by solving Eq.~(10) from \cite{Nieser:2011} (without integer constraints, without rounding). In this we use frame fields provided by the authors of \cite{liu2018singularity}, corresponding to the results shown in that article.
Numerical sanitization of these parametrizations (Sec.~\ref{sec:sanity}) took less than a second for most cases, 7s for the most complex case (with 163K tets).
Table~\ref{fig:statparam} shows details about these runs, including the number of facets traversed and forming the motorcycle complex walls.
Notice that again the base complex is significantly more complex; the number of blocks is up to $30\times$ higher, construction time up to $13\times$, memory consumption up to~$9\times$.

\noindent\emph{Remark:} The motorcycle complex is well-defined only for \emph{valid} seamless parametrizations in general.
 Their fully robust generation in 3D is a problem under broad investigation. In particular because our algorithm operates on generic \emph{continuous} rather than special \emph{quantized} parametrizations, it was easy, though, to yield valid input parametrizations for 15 out of 19 models from \cite{liu2018singularity} already with the above simple best-effort approach following \cite{Nieser:2011}.

\subsection{Example Use Cases}
\label{sec:example}

\subsubsection{Quantization for Hex Meshing}
\label{sec:quantization}

A \emph{continuous} seamless parametrization, as discussed in Sec.~\ref{sec:seamless}, can be viewed as defining an infinitely fine hexahedral mesh, whereas the special case of a \emph{quantized} seamless parametrization (also called integer-grid map) implies a finite hexahedral mesh.
In the 2D case, this relation is exploited in state-of-the-art quadrilateral mesh generation methods. \cite{Kaelberer:QuadCover} pioneered the idea of first generating a continuous seamless parametrization, and then \emph{rounding} it (at once or iteratively \cite{Bommes:MIQ}) to a quantized seamless parametrization.
This rounding is a notoriously fragile process, though: With increasing target quad size, the risk of yielding an invalid parametrization (with degenerate or flipped parts) increases, as pointed out and demonstrated in \cite[Fig.\! 1]{Bommes:RMIQ} and \cite[Fig.~3]{Campen:2015:QGP}. 
An analogous rounding procedure has been described for the 3D case \cite{Nieser:2011}; it is the state-of-the-art approach to yield quantized volumetric seamless parametrizations, as evidenced by its sustained use in recent works \cite{AMUCIP,solomon2017boundary,liu2018singularity,Corman:2019:SMF,Palmer:2020}. Not surprisingly it comes with the same limitations as its 2D counterpart, as also evidenced in Table~\ref{tab:statsroundvsmc}.

In the 2D case, subsequent work has provided a remedy, taking a different path, reliable and efficient, from continuous to quantized seamless parametrizations: via the motorcycle graph \cite{Campen:2015:QGP,Lyon:2019:PQF,Lyon:2021}; for the 3D case, this path has not been paved yet.
Our motorcycle complex is the key to extending this state-of-the-art approach to the 3D case, generating hexahedral meshes via volumetric seamless parametrizations.

As a proof of concept, to illustrate the potential, we translate a simple version of this approach to the 3D setting: we compute the motorcycle complex of a continuous seamless parametrization, and then
scale the parametrization within each block such that it adopts integer dimensions, trivially implying some $l~\times~m~\times~n$-grid of unit hexahedra per block. The dimensions and the scaling  need to be chosen such that these grids conform across block boundaries. This is achieved by expressing the quantization (the integer dimensions choice) by assigning integer lengths to arcs---shared between walls and blocks, inherently ensuring compatibility.

What we need to require for this assignment is that walls remain rectangles (thus blocks remain rectangular cuboids) parametrically. Let $A_i$, $i\in\{0,1,2,3\}$, denote the set of arcs forming the four sides of a wall, and $\ell_a \in \mathbb{Z}^{>0}$ the length assignment of arc $a$. Then this requirement can be expressed using two linear constraints per wall:
\vspace{-0.2cm}
\begin{equation}
\label{eq:opt:con}
    \sum_{a\in A_i}\ell_a = \sum_{a\in A_{i+2}}\ell_a, \quad i=0,1
\end{equation}
One aims to reproduce the sizing of the given parametrization using
\vspace{-0.2cm}
\begin{equation}
\label{eq:opt}
    \sum_{a}(\ell_a - s\|a\|)^2 \rightarrow \min,
\end{equation}
where 
$\|a\|$ denotes the original parametric length, and $s$ is a scaling factor that allows choosing the resulting mesh's resolution.

Note that the parametrization per block cannot be scaled by a simple affine map as each of the block's six facets consists of possibly multiple walls (due to T-joints from outside the block), and each wall needs to be scaled according to its arcs' values $\ell_a$. It can be achieved
via a \emph{piecewise}-affine map $\sigma$, though: affine per
\begin{wrapfigure}[5]{r}{0.265\linewidth}
\vspace{-3.1mm}
\hspace{-0.6cm}
\begin{overpic}[width=1.2\linewidth]{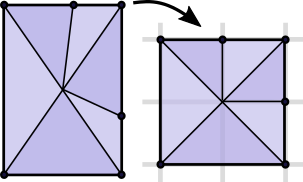}
\end{overpic}
\end{wrapfigure}
tetrahedron spanned by the block's center point with a triangle in a conforming triangulation of the (rectangular) walls on its surface. The inset illustrates a 2D version. If one splits the underlying tetrahedral mesh by these meta-tetrahedras' faces, $\sigma$ is affine per element and no inversions occur under $\sigma\circ\phi$ (most of this refinement is superfluous and can be omitted).
A smoothing of either the resulting parametrization \cite{SLIM} or the implied hex mesh \cite{PHOVER} can be applied subsequently to distribute distortion evenly.

\paragraph*{Comparison: Rounding}
To give an idea of the benefit, in Table~\ref{tab:statsroundvsmc} we compare this motorcycle complex based quantization strategy with the classical rounding strategy on a dataset of 19 tetrahedral meshes with frame fields, namely those shown in \cite{liu2018singularity}. 
It can be seen that for 15 of these a valid continuous seamless parametrization can be obtained by solving Eq.~(10) from \cite{Nieser:2011} to begin with---namely those used for the above experiments in Table~\ref{fig:statparam}. To these 15 continuous parametrizations we applied the rounding strategy, increasing the target edge length (via sizing $s$) until failure, and state the number of hexahedra implied by the coarsest rounded seamless parametrization that could validly be obtained. We also applied the motorcycle complex based quantization strategy, and state the number of hexahedra obtained when setting $s=0$ (aiming for maximal coarseness) in Eq.~\eqref{eq:opt}. It can be observed that, except for the simplest models, typically a significantly coarser quantization, thus coarser hex mesh can be obtained.

\begin{table}[t]
\vspace{-0.05cm}
\rowcolors{1}{}{backgray}
\begin{filecontents*}{tables/StatsMay7RoundingVsMC.csv}
Model;minHexesRound;minHexesMC;invalidPercent;invalidVolPercent;belowMinHexesRound;minHexPercent;minHexFactor
kitten;3112;176;9.1;1.135;broken;5.7\%;17.7
armadillo;30456;1884;7.3;1.035;broken;6.2\%;16.2
camille hand;1438;122;5.6;2.063;broken;8.5\%;11.8
bone;628;87;4.8;1.46;broken;13.9\%;7.2
sculpture;642;108;1;0.231;valid;16.8\%;5.9
sphere;32;7;5.3;4.14;valid;21.9\%;4.6
fandisk;502;110;1.3;1.09;valid;21.9\%;4.6
joint;888;205;1.6;0.901;broken;23.1\%;4.3
rockerarm;4784;1391;2.7;2.721;broken;29.1\%;3.4
prisma;9;3;2.3;1.543;valid;33.3\%;3.0
fanpart;13;5;0.6;0.295;valid;38.5\%;2.6
cylinder;60;26;3.6;1.54;valid;43.3\%;2.3
cube sphere;10;10;0;0;valid;100.0\%;1.0
tetrahedron;4;4;0;0;valid;100.0\%;1.0
broken bullet;36;44;0;0;valid;122.2\%;0.8
star bolt;-;-;0.1;0.028;valid;-;-
knot;-;-;0.9;0.892;broken;-;-
bunny;-;-;0.1;0.008;broken;-;-
elephant;-;-;0.2;0.002;broken;-;-
\end{filecontents*}
\csvreader[separator=semicolon, before reading=\begin{adjustbox}{max width=\columnwidth}, after reading=\end{adjustbox},tabular=lrrrrr, head to column names=true,
table head =
    \bfseries Model &
    \bfseries \;\;\;\;\;\;\;MC & 
    {\bfseries \;\;\;\;\;$\frac{\text{MC}}{\text{Round}}$} &
    \bfseries \;\;\;Round &
    \textcolor{gray}{\;\;\;\;inverted} &
    \bfseries \;\;\;\;HexEx\vspace{1mm}\\,
table foot =,
]{tables/StatsMay7RoundingVsMC.csv}{}
    {\textsc{\Model}\!\! &
    \minHexesMC & 
    \!\!{\minHexPercent} &
    \minHexesRound & 
    \textcolor{gray}{\invalidPercent\%} & 
    \belowMinHexesRound
    }
\vspace{0.0cm}
\caption{Statistics on the maximum coarseness (number of hexes) of hexahedral meshes generated from seamless parameterizations using our approach employing the motorcycle complex (MC), and via the 
classical method of iterative rounding (Round). We also report the percentage of parametrically inverted (or degenerate) tetrahedra when trying to achieve the coarseness of MC (or a very fine mesh in the four bottom rows) with the rounding-based approach. As can be seen in the last column, the HexEx approach from \cite{Lyon:2016:HexEx} is able to recover a valid hex mesh from these invalidly rounded parametrizations only in mild cases. Also see Fig.~\ref{fig:roundvsmc}.\label{tab:statsroundvsmc}}
\vspace{-0.2cm}
\end{table}

\begin{figure}[b]
\vspace{-0.25cm}
\centering
\includegraphics[width=\linewidth]{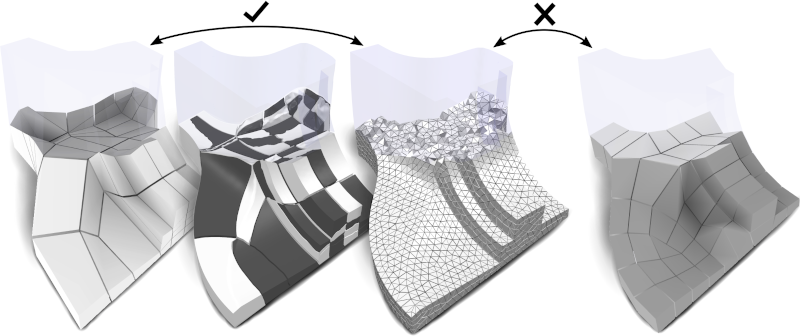}%
\vspace{-0.15cm}%
  \caption{Left: The quantization approach yields a hex mesh (left) with volumetric correspondence between each hex and a region (checkerboard piece) of the input object. Right: A hex mesh recovered by HexEx does not come with such a volumetric map.
  \label{fig:checker}}
\vspace{-0.1cm}
\end{figure}

\begin{figure*}[t]
  \centering
  \begin{overpic}[width=.99\linewidth]{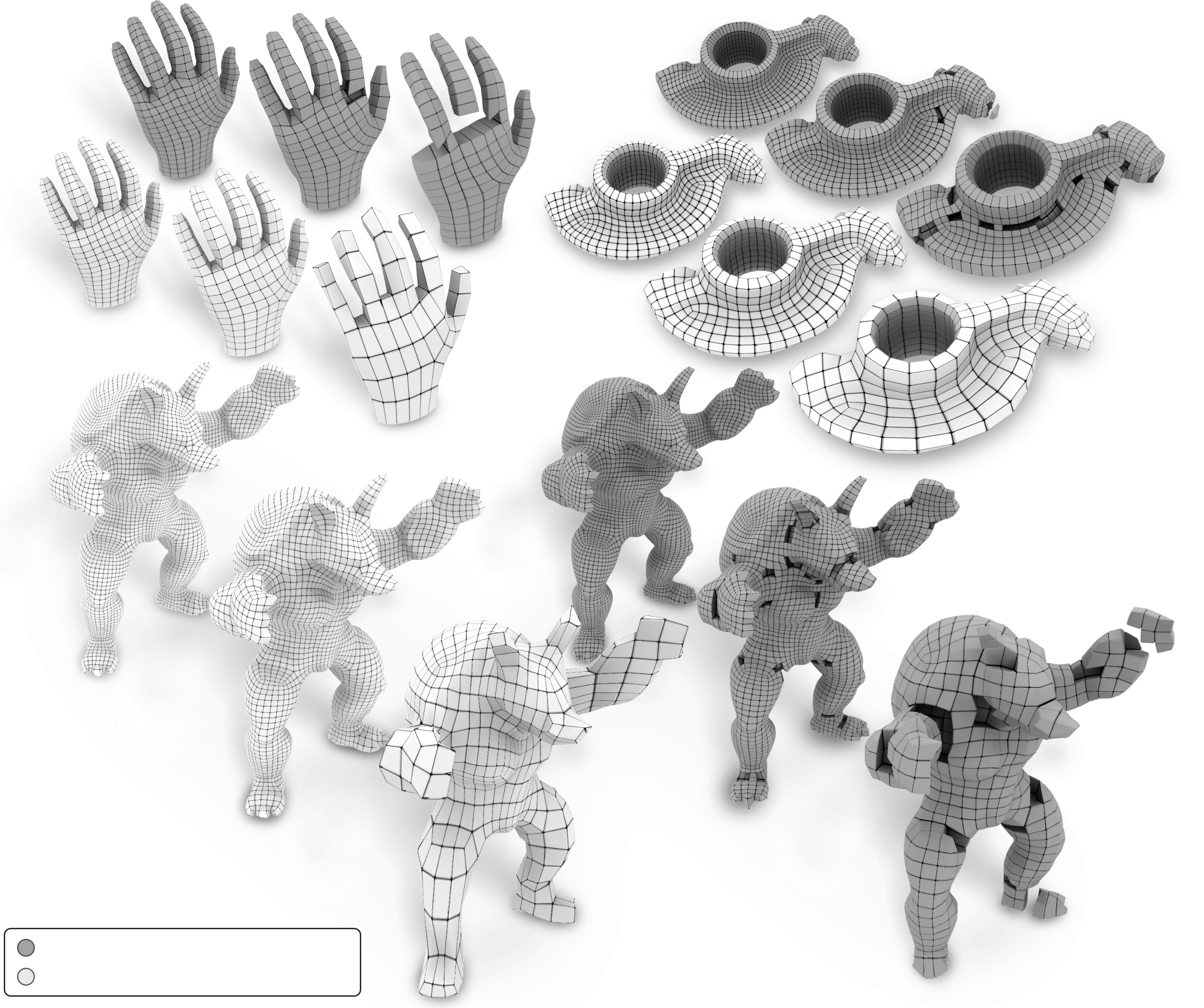}%
  \put(3.5,4.6){\small classical iterative rounding + HexEx}
  \put(3.5,2.2){\small motorcycle complex based quantization}
  \end{overpic}
  \vspace{-0.15cm}%
  \caption{Hexahedral meshes of increasing coarseness obtained from seamlessly parametrized tetrahedral meshes by means of iterative rounding to an integer-grid map as described by \cite{Nieser:2011} (dark gray) and our method using the motorcycle complex (white). When increasing target coarseness, hexahedral meshes obtained by rounding become increasingly defective due to parametric degenerations and inversions. Even fault-tolerant mesh extraction \cite{Lyon:2016:HexEx}, as employed here, cannot recover from this, leaving gaps that cannot easily be patched.
  \label{fig:roundvsmc}}
  \vspace{-0.25cm}%
\end{figure*}%

We remark that HexEx \cite{Lyon:2016:HexEx} can sometimes extract valid hex meshes even from invalid rounded parametrizations. We applied it to parametrizations rounded to the same level of coarseness as can be achieved with our approach. In {7} of the {16} cases where the rounded parametrization is invalid, it was able to output a valid hex mesh; in {9} cases the hex mesh has defects, some examples of which are shown in Fig.~\ref{fig:roundvsmc}. Note that when HexEx succeeds in ignoring the parametrization's defects, it does output a mesh but, in contrast to our approach, no valid parametrization, in particular no bijection between the input and the output mesh (Fig.~\ref{fig:checker}).

\begin{figure}[b]
\centering%
\vspace{-0.1cm}%
  \includegraphics[width=0.87\linewidth]{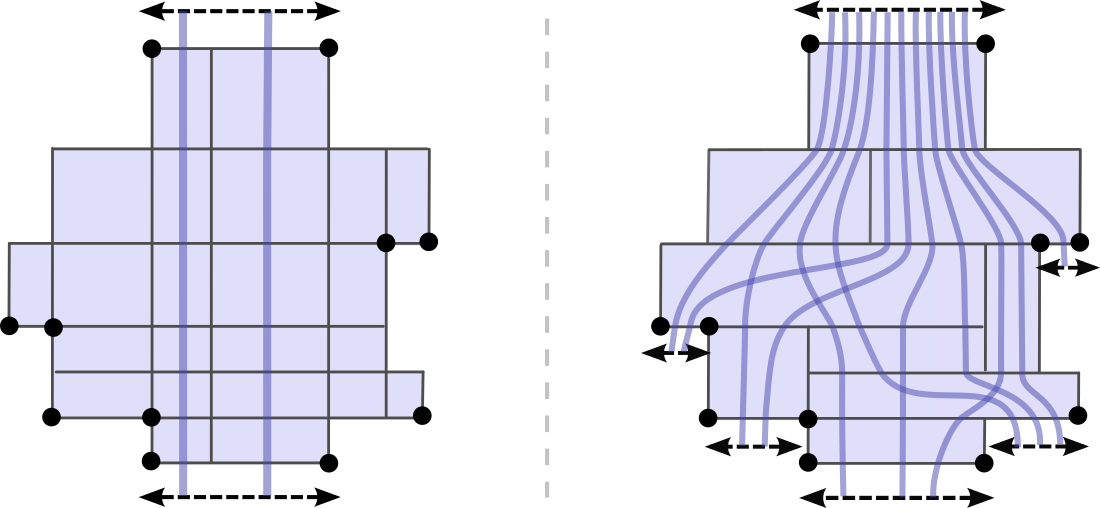}%
  \vspace{-0.05cm}
\caption{{Illustration of a part of a 2D slice through a BC (left) or MC (right); black dots are singularities. Assume the distance (parametric or number of hexes) between the upper two singularities shall be increased or decreased (in a quantization or a hex mesh). In the BC, this can be achieved using one of two \emph{sheet operators} (purple paths), here with essentially identical effect. In the MC, there are 11 different paths to adjust the quantization, with a choice of different effects on other singular, boundary, or feature points.}}\label{fig:sheetops}
\vspace{-0.15cm}
\end{figure}

\noindent\emph{Alternative: Coarsening.}
One may consider the alternative of only applying mild (therefore more robust) rounding, yielding an overly fine initial hex mesh, followed by coarsening, e.g., using \cite{Gao:2017}. Obvious downsides are the lack of a priori knowledge of a successful target edge length setting (potentially requiring trial-and-error) as well as the higher time and memory cost of this approach, operating fine-to-coarse rather than coarse-to-fine. Furthermore, the conforming sheet operators employed for structure and singularity preserving mesh coarsening are more restricted than a motorcycle complex based quantization procedure, cf.~Fig.~\ref{fig:sheetops}.

\noindent\emph{Alternative: Base Complex.}
For the same reason (in addition to complexity-related reasons), it is better to
formulate the quantization problem \eqref{eq:opt:con}+\eqref{eq:opt} based on the coarser and non-conforming motorcycle complex than on the base complex: there are more degrees of freedom, more ways for the solver to adjust the quantization length assignment $\ell$ (while respecting \eqref{eq:opt:con}), as illustrated in Fig.~\ref{fig:sheetops}. This in particular enables fine-grained control over the resulting mesh sizing (Fig.~\ref{fig:quantization}; also see supplemental material part~D).

\begin{figure}[t]
  \centering
  \vspace{-0.25cm}%
  \includegraphics[width=0.99\linewidth]{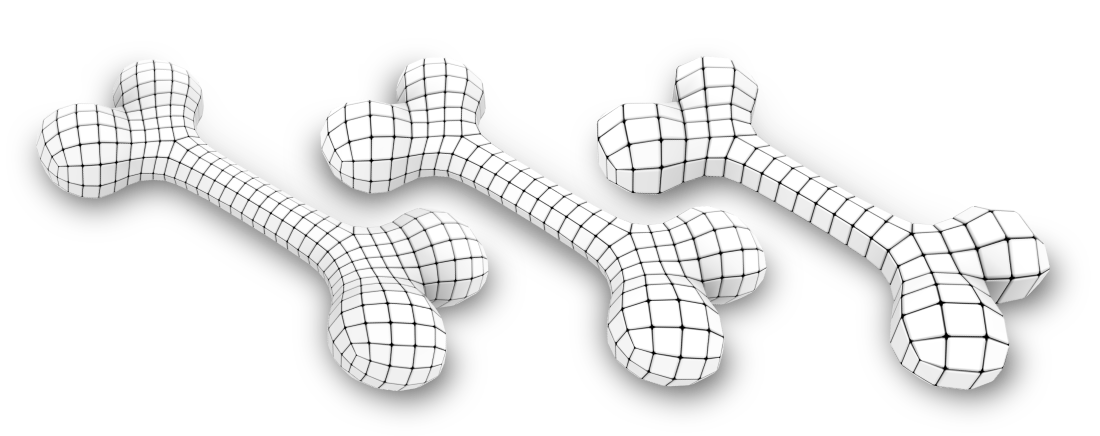}%
  \vspace{-0.38cm}%
  \caption{Hexahedral meshes of varying density obtained by quantizing a volumetric seamless parametrization using the proposed motorcycle complex. 
  \label{fig:quantization}}
 \vspace{-0.2cm}
\end{figure}

{We point out that, while these experiments already demonstrate various benefits, further improvements are possible and shall be explored in future work.}
For instance, we solve the integer program \eqref{eq:opt:con}+\eqref{eq:opt} using a general purpose solver (Gurobi); a tailored strategy, along the lines of \cite{Campen:2015:QGP}, could be more efficient.
Block reparametrization (to match the quantization) could be performed using efficient combinations of fast (e.g. discrete harmonic mapping) and reliable fallback (e.g. the piecewise-affine $\sigma$) solutions.
For simplicity, we required $\ell_a \!>\! 0$; supporting zero-arcs requires additional
efforts \cite{Lyon:2019:PQF} but will enable higher quality.

\subsubsection{Solid T-Splines}
\label{sec:spline}

\begin{figure}[b]
\vspace{-0.3cm}
\includegraphics[width=0.59\linewidth]{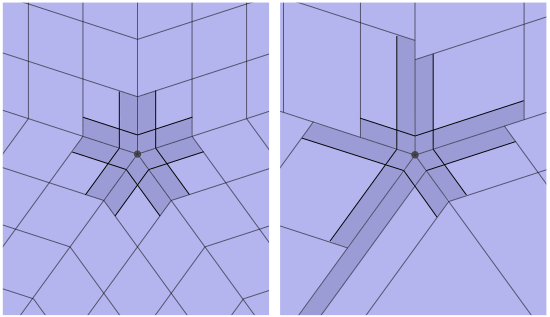}\phantom{.}
\rowcolors{1}{}{backgray}
\raisebox{0.89\height}{\resizebox{3.3cm}{!}{
\begin{filecontents*}{tables/tableTspline.csv}
Model;MCCrTsplineAdd;BCTsplineadd;factor
1;4190;12743;3.0
2;14656;28352;1.9
3;10736;14520;1.4
4;15407;31217;2.0
5;10608;17866;1.7
6;14167;26785;1.9
7;7348;14520;2.0
8;6437;11165;1.7
9;12720;24685;1.9
10;11303;19030;1.7
\end{filecontents*}
\csvreader[separator=semicolon, before reading=, after reading=,tabular=lrrr, head to column names=true,
table head =
\bfseries Model\!\!\! & \bfseries +BC & \bfseries +MC & \bfseries \!\!factor\vspace{0.3mm}\\,
table foot =,
]{tables/tableTspline.csv}{}
{\Model & \BCTsplineadd & \MCCrTsplineAdd & \factor$\times$}}}
\vspace{-0.52cm}
  \caption{
  T-spline-required refinement of blocks around a singular arc (cross section view) in a conforming (left) versus a non-conforming (e.g. MC) complex (center). Right: The T-mesh derived in this way from the MC is typically much coarser compared to the BC; on 10 example models from Table \ref{fig:stats}, the number of additional refinement-induced cells (+BC and +MC, respectively) is 1.4-3.0 times larger for the (already initially larger) BC.}\label{fig:spline}
\vspace{-0.15cm}
\end{figure}

T-splines are a flexible tool in the context of smooth function representation, for geometric modelling as well as for isogeometric analysis.
For the volumetric case, constructions of T-spline spaces starting from hexahedral meshes have been described \cite{Zhang:2012b}.

As solid T-splines are defined over cuboid complexes which are not necessarily conforming (hence the `T'), it is actually unnecessarily restrictive to start from a conforming one, i.e., a hexahedral mesh. We can essentially apply the necessary structural refinement around singularities that the above paper describes directly on the non-conforming motorcycle complex. This circumvents the need for quantization (to obtain a hex mesh), and effectively provides a coarser starting configuration---which could then be adaptively refined where necessary for a particular application, as opposed to starting with a rather dense hexahedral mesh as domain structure and later coarsening it where possible.

In Fig.~\ref{fig:spline} we illustrate the refinement around a singularity (inserting additional walls) by the rules of \cite{ZHANG2012185}, so as to yield a T-mesh suitable as control mesh for a solid T-spline. 
We note that some additional modifications to the control mesh structure can be necessary to ensure continuity due to non-local overlaps of basis function supports with singularities, as discussed in \cite[\S8.2]{campen2017similarity}, or to yield specific classes of splines, such as analysis-suitable splines, as discussed in \cite{scott2012local}. In any case, the motorcycle complex provides a significantly simpler starting point than the base complex (or a hexahedral mesh derived from it). We contrast the numbers of additional T-mesh blocks due to refinement-at-singularities applied to the BC and the MC in Fig~\ref{fig:spline}.

\vspace{-0.18cm}

%% file: text/08conclusion.tex
\section{Conclusion}

We have introduced a generalization of the motorcycle graph to the volumetric setting, providing a structure and algorithm that enable the compact block decomposition of solids. Hexahedral meshes or seamless volume parametrizations can serve as underlying basis.
We expect the 3D motorcycle complex will enable progress in various ways, just like the 2D original has found effective use in the context of parametrization, mesh generation, and mesh processing, in particular when it comes to providing robustness guarantees. We have demonstrated that our generalization has the potential to form the basis for extensions of such techniques to the (even more relevant) 3D cases where such guarantees are still lacking.

\vspace{-0.15cm}
\subsection*{Limitations \& Future Work}

While we have demonstrated that the motorcycle complex typically is very small, it is not necessarily the smallest non-conforming cuboid partition. Finding the actual minimum partition is very hard already in the 2D case \cite{eppstein2008motorcycle}. It would be interesting (even if not necessarily of high practical relevance) to investigate whether the size of the motorcycle complex relative to the minimal size is, as in 2D, bounded in some nice manner.

{In our prototype implementation we employ a very generic polyhedral mesh data structure (OpenVolumeMesh). In this case the computation 
of the complex is dominated by the tetrahedral splits (around 0.3ms per split on a commodity PC). A tailored lightweight data structure could likely reduce this.}

We have demonstrated the use case of quantization (Sec.~\ref{sec:quantization}), where the motorcycle complex can serve as key ingredient in the process of hexahedral mesh generation. Further developments in this direction, e.g., additionally enabling zero-quantizations for coarse meshes,
specializing solvers for efficiency, are of high relevance for ongoing developments in the field of mesh generation.

Some form of generalization to hex-dominant meshes, analogous to the 2D case \cite{schertler2018generalized}, could furthermore be of interest. This comes with additional challenges due to the greater structural variability compared to quad-dominant meshes.
For the parametrization-based algorithm, the incorporation of some form of tolerance to defects (degeneracies, local inversions) would broaden practical applicability. This direction has not even been explored for the 2D case yet, but insights from fault-tolerant mesh extraction \cite{ebke2013qex,Lyon:2016:HexEx} may provide inspiration.

%% file: text/99backmatter.tex
\printbibliography
\balance